\documentclass[11pt]{article}
\usepackage[pdfmenubar=false, letterpaper=false, pdfpagelabels=true]{hyperref}
\usepackage{amsmath}
\usepackage{fancybox}
\usepackage{epsfig}

\newcommand{\bm}[1]{\mbox{\boldmath$#1$}}
\def\mvec#1{{\bm{#1}}}   

\begin{document}
\title{Improved iterative Bayesian unfolding} 
\author{G. D'Agostini
\mbox{} \\
{\small Universit\`a ``La Sapienza'' and INFN, Rome, Italy} \\
{\small (giulio.dagostini@roma1.infn.it, 
\url{http://www.roma1.infn.it/~dagos})}
}
\date{}

\maketitle

\begin{abstract}
This paper reviews the basic ideas behind 
a Bayesian unfolding published some years ago 
and improves their implementation. In particular,
uncertainties are now treated at all levels by 
probability density functions and their propagation
is performed by Monte Carlo integration.
Thus, small numbers are better handled and 
the final uncertainty does not rely on the assumption
of normality. 
Theoretical and practical issues concerning the
iterative use of the algorithm are also discussed. 
The new program, implemented in the R language, is freely available,
together with sample scripts to play with toy models. 
\end{abstract}
\vspace{1.0cm}

\section{Introduction}
Physicists like to think of `true' distributions of physics
quantities, i.e those distributions 
(graphically represented by histograms and here also referred as `spectra') 
one would observe under idealized conditions
that seldom -- never, strictly speaking --  happen in real live
(ideal detector, no physical or instrumental background). 
The `observed' distribution is then considered as a 
'noisy distortion' of the true one. 
An important task of the experimental method is 
therefore to infer the true distribution from the observed one, i.e.
to correct the observed spectrum for distortion and noise. 
This can be done by  different methods that follow
different approaches.
Since this is not a review paper, 
I just outline the two different classes of strategies and then 
focus on the specific issues of this work. 

In a first kind of approach a mathematical function for 
the true distribution
is assumed (together with other functions to model the noise)
and the task becomes that of 
estimating the free parameters of the function(s). Indeed, we
fall in the so called domain of {\it parametric inference} (`fits'), 
usually associated to names like `least-squares' or 
'maximum likelihood'.\footnote{However one can show that it is preferable
to base the data analysis on more 
solid probabilistic ground,
from which the mentioned methods might be derived
(see e.g. Ref.~\cite{lfits}) under special  conditions 
which fortunately hold in the large majority of 
practical cases (but it is better one  knows 
when the conditions of validity hold 
and what to do when they do not!).}  

In parametric inference all information contained in the observed 
spectrum is, to say, `distilled' into the model parameters.
Parametric inference, especially when the 
conditions to use least-squares methods
hold, is usually the best and fastest way to proceed, 
if we have good reasons
to believe the hypothesized family of functions.

However, sometimes we wish
to interpret the data as little as possible and just public 
`something equivalent' to an experimental distribution, 
with the bin contents fluctuating according to an underlying 
multinomial distribution, but having possibly got rid of physical
and instrumental distortions, as well as of background.
In particle physics this second approach goes 
under the name of {\it unfolding}
(or deconvolution, or restoration). 

Several years ago a simple algorithm 
based on   Bayes' theorem was presented ~\cite{BU}
with which it was possible
to achieve rather `good' results (`good'
compared with the difficulty of the task). 
The main improvements presented here concern the handling 
of small numbers and the evaluation of the uncertainty on the 
unfolded distribution, while the guiding ideas 
and the basic assumptions are substantially unchanged. 
To be more clear, 
the algorithm of Ref.~\cite{BU} was relying on 
the underlying hypotheses of normality and `small relative errors':
`best estimates' were provided, with uncertainty calculated
from standard `error propagation' 
formulas.\footnote{It is worth remembering that these formulas
rely on an hypothesis of linearity between {\it input} and 
{\it output quantities}, an hypothesis that holds approximately, 
in non-strictly linear problems, if the relative uncertainties of 
the input quantities are small enough, such that the 
dependence input$\rightarrow$output can be locally
linearized~\cite{asymmetric}.} 
The new algorithm handles better small numbers, 
in the way it will be described in Sec.~\ref{ss:improvements},
and performs the propagation of uncertainty by sampling,
i.e. by Monte Carlo (MC) integration.
The algorithm has been implemented in a R language~\cite{R} code available 
on the author's web page~\cite{R-code}.

The paper is organized in the following way. 
Section \ref{sec:BayesIntro} gives a short introduction to
Bayesian inference and to the specific application subject of this paper. 
Then the algorithm of  Ref.~\cite{BU} is reminded 
and the improvements are presented.
The issue of iterative use of the algorithm
is also discussed, although the program now gives  also the possibility
to use priors, provided by the user, over all possible true
spectra. 
This option should avoid the
need of iterations (but I anticipate that it might be  not that easy to 
model such priors and then the iterative strategy
remains a pragmatic solution).
Finally, some results on toy models are presented. 

Some technical issues concerning binomial and Dirichlet
distribution, including the use of the latter as prior
conjugate of the former, are reminded in Appendix A. 
A second appendix is dedicated to the handling of the 
zeros occurring in the evaluation of the smearing matrix
(a cause that produces no event in some effect bins, as it 
usually happens, depending also on the statistics of the
Monte Carlo simulation). As it happens with intermediate 
smoothing (or other kinds of regularization), this is
a suggestion of how the problem can be approached, whose
solution is delegated to the user, who is supposed to
know the physics case under investigation.

\section{Bayesian inference and Bayesian unfolding: \\
from first principles to real life}
\label{sec:BayesIntro}
The so called Bayesian inference is a way to learn 
about physical quantities and their relationships 
(all things we cannot directly see) 
from experimental data (what we can actually observe 
with our senses, usually mediated by more or
less sophisticated detectors)
using probability theory. This game
is {\it conceptually} rather
simple, proviso we accept that the intuitive meaning 
of probability -- a scale to rank our beliefs that 
several events might happen, or that several hypotheses 
might be true -- is suitable in scientific reasoning
(see Refs.~\cite{BR} and \cite{rpp} and references therein
for an introduction). 

\begin{figure}[!t]
\begin{center}
\begin{tabular}{cc}
\epsfig{file=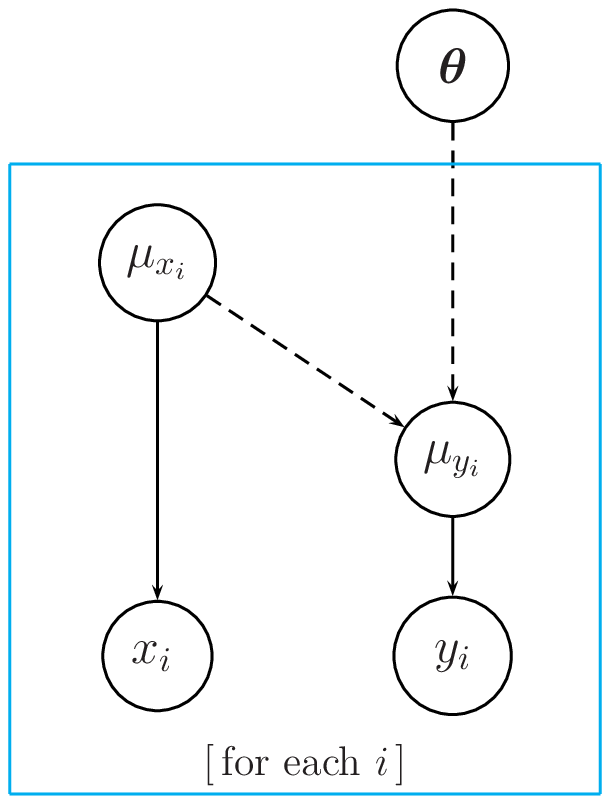,clip=,width=0.37\linewidth} &
\epsfig{file=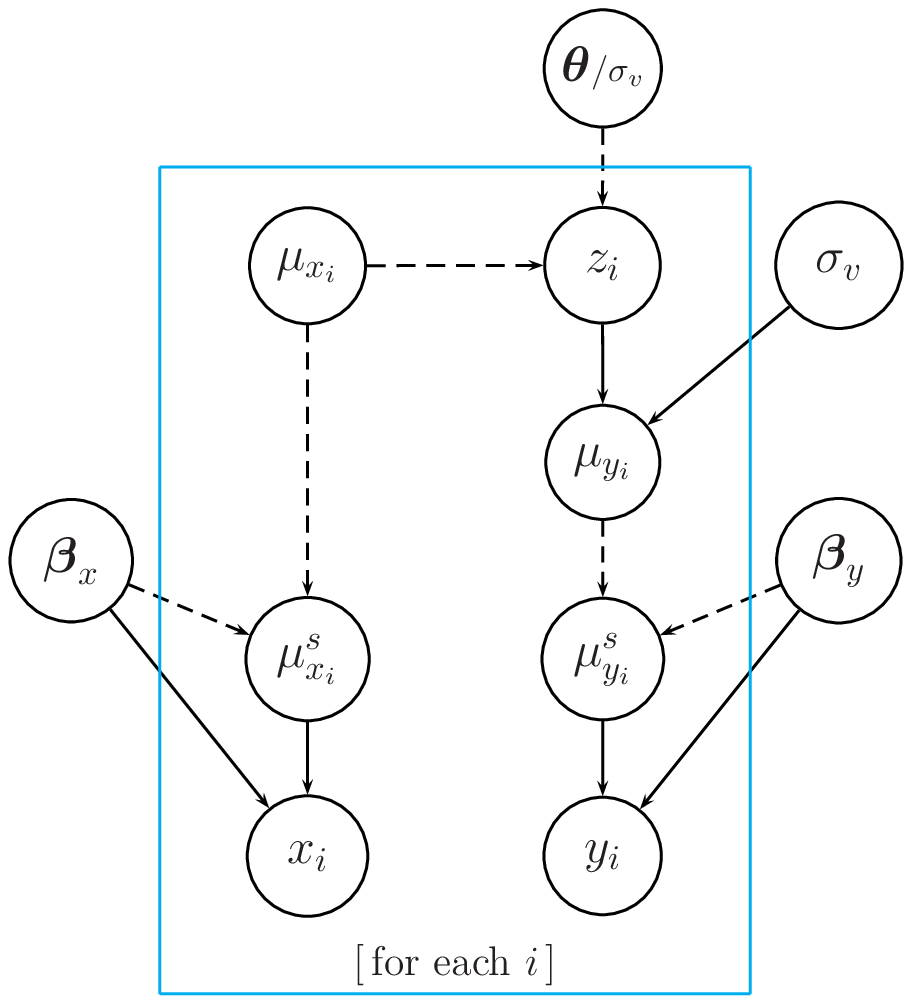,clip=,width=0.55\linewidth} \\
{\large {\it a)}} & {\large {\it b)}}
\end{tabular}
\end{center}
\caption{\sl Bayesian network describing a fit model~\cite{lfits}. 
$x_i$ and $y_i$ 
are the experimental observations, related to  $\mu_{x_i}$ and $\mu_{y_i}$
by experimental errors. Instead, a deterministic `law' connects 
the `true' values 
$\mu_{y_i}$ to  $\mu_{x_i}$ via the model parameters  $\mvec\theta$
(solid arrows stand for probabilistic links, dashed for deterministic).
Network {\it a)} describes a simple model with errors only on 
the ordinate. Network {\it b)} takes into account errors on both
axes, extra variability of the data points around the believed
physical law and systematic effects.}
\label{fig:bn1}
\end{figure}
The starting point of a Bayesian inference is to build up a model 
for the deterministic 
(``{\it B follows from A}'')
and  the probabilistic 
(``{\it B might follow from A}'')
connections
that relate the several entities that enter the problem.
This model has the interesting graphical representation
of a network, usually called {\it Bayesian network} or 
{\it belief network} 
(to get an idea of their meaning and their application 
{\it Google} these keywords, or browse the 
\href{http://en.wikipedia.org/wiki/Bayesian_network}{\it Wikipedia}). 
For example, Fig.~\ref{fig:bn1}, taken from 
Ref.~\cite{lfits}, shows a Bayesian network
to model two physical quantities, $\mu_x$ and $\mu_y$,
connected to each other with a `law', whose parameters 
are denoted by  $\mvec\theta$. In the 
very elementary case depicted in the 
 left hand network of Fig.~\ref{fig:bn1}
 the solution to the problem can be rather simple, 
under some assumptions that, fortunately, hold
very often  in routine cases.
But, in general, the solution is not 
that simple. Nevertheless, as stressed in Ref.~\cite{lfits},
after one has built the graphical model, one is 
often more than half the way to the solution, 
thanks to the great progresses  recently made in 
Bayesian network computing.

\begin{figure}
\begin{center}
\epsfig{file=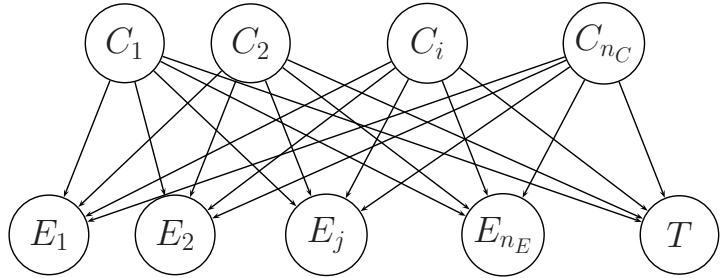,clip=,width=0.75\linewidth}
\end{center}
\caption{{\sl Probabilistic links from causes to effects. 
The node indicated by $T$ (`trash') stand for the 
inefficiency bin and corresponds to $E_{n_E+1}$}}
\label{fig:unfnet}
\end{figure}
The essence of the Bayesian unfolding of Ref.~\cite{BU},
that is the starting point also of this new version,
is to make the problem discrete and to treat the `cause'
bins as independent degrees of freedom, i.e. without 
constraints among each other.\footnote{An important 
drawback of this feature will be discussed in section \ref{sec:iter}.}
For this reason the algorithm
can virtually handle any kind of `smearing' and it is
easily extensible, 
at least in principle and with the only limit due to computer power, 
 to multidimensional
problems. In fact, the core of the algorithm only knows about 
`cause-cells' and `effect-cells',\footnote{`Bin' 
and `cell' are used here as synonyms, although `cell' refers
to a region on the configuration space and 
`bin' to histograms representing them.} 
but it does not know the 
location of the cells in the configuration space of the problem.  
For the same reason, the treatment of background, 
and even of several independent sources of background, 
can be easily embodied in the algorithm
by just adding extra cause-cells, one cell per source of background.   
As a by-product, the algorithm also provides the number of events
to be attributed to each source of background.
(It is worth remembering that 
background might have an interesting physical meaning,
and thus the estimation of the level `noise' 
might provides indeed a physics measurement, 
as in the analysis of Ref.~\cite{Raso2}.)

Given the discretization of the problem,
the Bayesian network relating 
causes and effects is that shown in Fig.~\ref{fig:unfnet},
where we use the same notation of Ref.~\cite{BU}, with the addition 
of the effect bin $T$ (`trash'), equivalent  to  $E_{n_E+1}$, 
to describe inefficiency (the reason to introduce this 
extra bin will become clear later).

\begin{figure}[!t]
\begin{center}
\vspace{0.7cm}
\begin{tabular}{c}
\epsfig{file=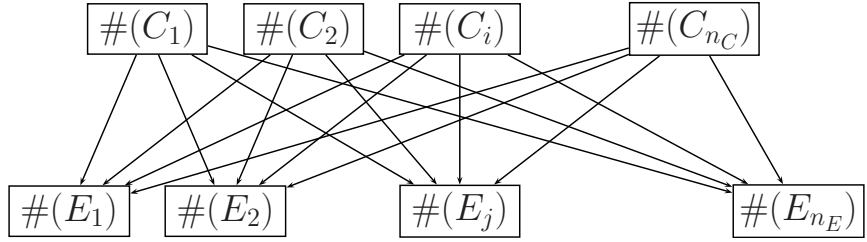,clip=,width=0.9\linewidth} 
\end{tabular}
\end{center}
\caption{{\small From the `true' distribution (numbers of events in the 
cause-bins) 
to the observed distribution (numbers of events in the 
effect-bins). The number of events $\#(\,)$ is indicated with 
$x(\,)$ in the text.}}
\label{fig:bn0}
\end{figure}

Rephrasing the problem in probabilistic terms, the purpose of the 
unfolding is to find the `true' number of events  
in each cause bin [$\#(C_i)$ in Fig.~\ref{fig:bn0},
indicated by $x(C_i)$ in the text],
given the observed spectrum and assuming 
some knowledge about the smearing.
  
Since the links cause$\rightarrow$effects have a probabilistic nature,
it follows that also
the links effect$\rightarrow$causes will be probabilistic, 
and therefore it will be
uncertain  the number of events
to be attributed to the cause-cells. We can only attempt to rank in 
probability all possible spectra that might have caused the observed
one. 
In other words, the realistic goal of our analysis
is not to determine the {\it the} true spectrum, but rather to assess
\begin{eqnarray}
P(\mvec x_C \,|\,\mvec x_E,\,\Lambda,\,I)\,,
\label{eq:goal}
\end{eqnarray}
where:\footnote{In Ref.~\cite{BU} 
$\mvec x_C$, $\mvec x_E$ and $\Lambda$ were respectively indicated
by  $\underline{n}_C$, $\underline{n}_E$ and $\mvec S$.
As in Fig.~\ref{fig:bn0} and in Ref.~\cite{BU}, the symbols $i$
and $j$ are used to index causes and effects, respectively, 
no matter if this choice leads to the unusual convention of 
indexing the $\Lambda$ rows by $j$ and the columns by $i$, as in 
Eq.~(\ref{eq:Lambda}). 
Note that, when we refers to MC simulations
to infer the smearing matrix, $\mvec x_E$, we
 also need include the trash bin, as it will be reminded
at the proper place.
Later on [see Eq.~(\ref{eq:lambda_i})] it will be convenient
to name $\mvec{\lambda}_i$ the columns of the matrix $\Lambda$. 
In summary
\begin{eqnarray}
\Lambda &=& \left(\!\!
            \begin{array}{cccc}
             P(E_1\,|\,C_1,\,I) &  P(E_1\,|\,C_2,\,I) & \ldots &  P(E_1\,|\,C_{n_c},\,I) \\
             P(E_2\,|\,C_1,\,I) &  P(E_2\,|\,C_2,\,I) & \ldots &  P(E_2\,|\,C_{n_c},\,I) \\ 
             \ldots &  \ldots &  \ldots &  \ldots  \\
             P(E_{n_E+1}\,|\,C_1,\,I) &  P(E_{n_E+1}\,|\,C_2,\,I) 
             & \ldots &  P(E_{n_E+1}\,|\,C_{n_c},\,I)
            \end{array}
            \!\!\right) \nonumber \\
        &=& (\mvec{\lambda}_1,\,\mvec{\lambda}_2,
                          \ldots,\mvec{\lambda}_{n_C})\,. \nonumber
\end{eqnarray}
}
\begin{itemize}
\item
$\mvec x_C = \{x(C_1),\, x(C_2),\, \ldots,\, x(C_{n_C}) \}$ 
is the number of events in each bin of the true distribution,
i.e. {\it a} true spectrum.\footnote{For 
the use of the indefinite article in conjunction with 
`true values', see the ISO 
{\it ``Guide to the expression of uncertainty
in measurement''}\,\cite{ISO}, 
according to which {\it a} true value is
{\it ``a value compatible with the definition
of  a given particular quantity.''} }
\item
$\mvec x_E = \{x(E_1),\, x(E_2),\, \ldots,\, x(E_{n_E}) \}$
is the observed spectrum. 
\item
$\Lambda$ stands for the smearing matrix, whose elements 
$\lambda_{ji}$ (see remarks in footnote 4 concerning notation) are 
defined in probabilistic terms as
\begin{eqnarray}
\lambda_{ji} &\equiv& P(E_j\,|\,C_i,\,I)\,. 
\label{eq:Lambda}
\end{eqnarray}
The knowledge of $\Lambda$ comes usually from MC simulation and it is
therefore affected by an uncertainty, described,
 in general terms, by a pdf $f(\Lambda\,|\,I)$.
\item
$I$ stands for the state of information 
under which the analysis is performed
(this underlying condition  is often implicit in the pdf's).
\end{itemize}
Once we have stated clearly and in probabilistic terms our question
({\it  ``what is $P(\mvec x_C \,|\,\mvec x_E,\,I)$?''}),
probability theory provides us the answer,\footnote{Instead, 
the idea of inverting the smearing matrix (assuming it square
and not singular) is wrong in principle 
(i.e. besides the ascertainment 
 that such a method  yields unacceptable results). 
In fact, unfolding
is a probabilistic problem and not a 
deterministic linear one (or, equivalently, a geometric problem 
of rotating vectors). Therefore it needs to be solved by probabilistic
tools and not by linear algebra methods (`rotations').
 It is easy to show that
`rotation' works only when we 
have an `infinite' number of events, such that stochastic
effects are negligible and the observations coincides with the 
expectations. In fact, each product $\lambda_{ji}\,x(C_i)$ is nothing but 
the expected number of events in the effect-cell $E_j$ due to cause 
$C_i$ alone: 
\begin{eqnarray*}
\mbox{E}[\left.x(E_j)\right|_{x(C_i)}] &=& P(E_j\,|\,C_i,\,I)\,x(C_i) =
           \lambda_{ji}\,x(C_i) 
\end{eqnarray*}
Summing up the contributions from all cause-cells, we get the
expected value in the effect-cell $E_j$
\begin{eqnarray*}
\mbox{E}[\left.x(E_j)\right|_{\mvec x_C}] &=& 
 \sum_i \lambda_{ji}\,x(C_i)\,,
\end{eqnarray*}
that can be written in matrix form as
\begin{eqnarray*}
\mvec \mu_E \equiv \mbox{E}[\mvec x_E] &=& \Lambda \, \mvec x_C\,.
\end{eqnarray*}
Then, if $\Lambda$ is square and not singular, we get
\begin{eqnarray*}
\mvec x_C &=& \Lambda^{-1}\,\mvec \mu_E \,.
\end{eqnarray*}
But this might be, at most, the solution of a text book exercise in 
mathematical probability theory, and does not help
to solve real problems. 
This is the reason why, besides
the fact that the matrix inversion gives notoriously 
bad results,
{\it the very idea 
of inverting the smearing matrix is logically
flawed}: we can certainly apply $\Lambda^{-1}$ to 
a vector of numbers {\it already known} to be sums 
of expected values of binomials, but we cannot apply
it to a vector of numbers that {\it might be} 
(we are not even sure of this, because some counts could be
due to background we do not take into account!) 
sums of binomial random variables. If we do it, there is no
guarantee that  $\Lambda^{-1}\mvec x_E$ yields a vector
of `valid numbers' of the $n$-parameters of binomials 
(the question that they might have the meaning of 
a physical spectrum for the problem under study is 
secondary at this level) and, in fact, even negative numbers
can be obtained!
It follows that unfolding methods which
use the matrix inversion as starting point and try to
cure its bad features with some kitchen are not appealing from 
a theoretical point view, although they might even provide 
acceptable results because of `mysterious' reasons
I do not want to enter into (cooks might be extremely clever!).
}
at least in principle. 
In fact, 
\begin{enumerate}
\item
Bayes' theorem allows to calculate 
$P(\mvec x_C \,|\,\mvec x_E,\,\Lambda,\,I)$, given the observation 
$\mvec x_E$ and the smearing matrix $\Lambda$, as
\begin{eqnarray}
P(\mvec x_C \,|\,\mvec x_E,\,\Lambda,\,I) &=& 
\frac{P(\mvec x_E \,|\,\mvec x_C,\,\Lambda,\,I) \cdot P(\mvec x_C\,|\,I)}
     {\sum_{\mvec x_C}P(\mvec x_E \,|\,\mvec x_C,\,\Lambda,\,I) 
     \cdot P(\mvec x_C\,|\,I)}
\label{eq:bayes1}
\end{eqnarray}
(the formula will be explained in a while).
\item
We can take into account of the uncertainty about $\Lambda$
using another theorem of probability theory, namely
\begin{eqnarray}
P(\mvec x_C \,|\,\mvec x_E,I) &=& \int 
P(\mvec x_C \,|\,\mvec x_E,\Lambda,\,I)\,
f(\Lambda\,|\,I)\,\mbox{d}\Lambda\,.
\label{eq:marg_lambda}
\end{eqnarray}
\end{enumerate}
Let us now go through the several ingredients needed to get 
$P(\mvec x_C \,|\,\mvec x_E,I)$
and try to understand 
were the problems arise. 
\begin{itemize}
\item
First of all, it is easy to realize that the denominator of 
Eq.~(\ref{eq:bayes1}) is just a normalization factor, and then 
we can rewrite the Bayes' formula in a way that 
highlights the main ingredients:
\begin{eqnarray}
P(\mvec x_C \,|\,\mvec x_E,\,\Lambda,\,I) & \propto & 
P(\mvec x_E \,|\,\mvec x_C,\,\Lambda,\,I) \cdot P(\mvec x_C\,|\,I)\,,
\label{eq:bayes2}
\end{eqnarray}
where $P(\mvec x_E \,|\,\mvec x_C,\,\Lambda,\,I)$ is the 
so called {\it likelihood}
 and $P(\mvec x_C\,|\,I)$ the 
 {\it prior}. The left hand side of the Bayes' formula takes
the name of {\it posterior}. As we see, 
in probabilistic inference the
likelihoods have the role of updating probabilities. 
\item 
Although the presence of priors in the formula might
cause anxiety in those who approach this kind
of reasoning for the first time,
 it is a matter of fact that: 1) priors are logically necessary 
to get $P(\mvec x_C \,|\,\mvec x_E,\,\Lambda,\,I)$
starting from the likelihood, i.e. to 
perform the so-called `probability inversion';
2) they allow to plug into the model all relevant prior information,
that might come from previous data or from theoretical
prejudices; 
3) priors are often so vague (or there are so many data -- that 
is the same for the relative balance of prior and likelihood 
in the Bayes' formula) that they influence negligibly the posterior,
and the inference is often dominated by the likelihood. 
\item 
Let us assume this last case of prior vagueness applies. 
This is equivalent to have 
\begin{eqnarray}
P(\mvec x_C\,|\,I) & = & \mbox{\it constant}
\label{eq:flat_prior}
\end{eqnarray}
and the  inference is then performed according to the rule
\begin{eqnarray}
P(\mvec x_C \,|\,\mvec x_E,\,\Lambda,\,I) & \propto & 
P(\mvec x_E \,|\,\mvec x_C,\,\Lambda,\,I)\,.
\label{eq:bayes3}
\end{eqnarray}
It is then not a surprise that the most probable spectrum  
$\mvec x_C$ is the one which maximizes the likelihood, 
{\it if} we have no other good reason to believe 
that this is not the case.
(This is the meaning of the expression ``recovering 
maximum likelihood estimators'' from the Bayesian approach.)   
\item
If we were able to write down a closed expression for the 
likelihood, or at least to provide a simple 
algorithmic expression of it,
we could somehow scan all possible spectra $\mvec x_C$,
find the most probable one (or an `average' spectrum that
summarizes in some sense the variety of true spectra 
compatible with the data)
 and assess somehow the uncertainty 
about the result. 
%
%
But, unfortunately, this is not the case
with $P(\mvec x_E \,|\,\mvec x_C,\,\Lambda,\,I)$ 
of our interest. 
Let us see why.
Given a certain number of events in a cause-bin $x(C_i)$, 
the number of events in the effect-bins,
included the `trash' one, is described by a 
{\it multinomial distribution} (see Appendix A.1):
\begin{eqnarray}
P(\mvec x_E\,|\,x(C_i), \Lambda, I) &=& 
\frac{x(C_i)!}{\prod_j^{n_E+1} x(E_j)!}\, 
\prod_j^{n_E+1}  \lambda_{ji}^{x(E_j)}\,.
\end{eqnarray}
It follows that $P(\mvec x_E \,|\,\mvec x_C,\,\Lambda,\,I)$ 
is the sum of  independent  multinomial distributions, 
for which, unfortunately, a closed formula does not 
exist.\footnote{It is well know that the sum of two Poisson variables
is still a Poissonian. Similarly, the sum of two binomial variables 
having {\it the same} parameter `$p$' is still a binomial. 
(This `reproductive property' applies to few other distributions). 
But the sum of two binomial variables with different $p$ 
does not have a closed expression 
(this can be understood either intuitively, just thinking to the
meaning of a binomial, or, more formally, analyzing
the product of the characteristic functions of each binomial). 
The same happens with a multinomial,
that is just an extension of the binomial to more than two 
possible outcomes.} This is the real serious 
technical problem that prevents
a straight application of the Bayes' formula.
\item
The elements of the smearing matrix $\Lambda$ are obtained
by MC: we generate a large number of events in each cell $C_i$ 
and count `where they end' after a realistic simulation. 
Intuitively, we expect $\lambda_{ji}\approx x(E_j)^{MC}/x(C_i)^{MC}$, 
but we also know  that this is just an estimate, with some 
uncertainty. Fortunately, in this case
we can make direct use of the Bayes' formula
applied to MC events, if we model the prior using 
a convenient pdf. 
In fact, if we indicate by $\mvec \lambda_i$ the 
$i$-th column of $\Lambda$ (see footnote 4), i.e. 
\begin{eqnarray}
\mvec \lambda_i = \{\lambda_{1,i},\,\lambda_{2,i},\ldots,\, 
\lambda_{n_E+1,i}\}\,, \label{eq:lambda_i}
\end{eqnarray}
we have
\begin{eqnarray}
f[\mvec \lambda_i\,|\, \mvec x_E^{MC},\,x(C_i)^{MC},\, I] 
&\propto& P[\mvec x_E^{MC}\,|\,x(C_i)^{MC}, \, \mvec \lambda_i,\, I]\cdot 
f(\mvec \lambda_i\,|\,I)\,. \nonumber \\
&& 
\end{eqnarray}
Since $ P[\mvec x_E^{MC}\,|\,x(C_i)^{MC}, \, \mvec \lambda_i,\, I]$
is a multinomial,
choosing a Dirichlet prior
(an extension of the Beta distribution to many dimensions 
-- see Appendix A.2),
we get a  Dirichlet posterior (Appendix A.3), i.e.
\begin{eqnarray} 
\mvec \lambda_i & \sim & \mbox{Dir}(\mvec\alpha_{posterior_i})\,,
\end{eqnarray}
with
\begin{eqnarray}
\mvec\alpha_{posterior_i} &=& \mvec\alpha_{prior_i} + 
\left.\mvec x_E^{MC}\right|_{x(C_i)}\,.\label{eq:update_alpha}
\end{eqnarray}
(Hereafter the symbols `$\sim$' stands for `follows the probability
distribution`.)
\item
Finally there is the integral (\ref{eq:marg_lambda}), which
is in principle not a problem 
(for example, thanks to modern technologies, it can easily performed 
by MC methods). 
\end{itemize}
In conclusion, the main serious problem is 
$P(\mvec x_C \,|\,\mvec x_E,\Lambda,\,I)$. Thus
being unable to tackle the problem from the main door, 
we need some `tricks' to circumvent the obstacle, but still under
the guidance of probability theory.

\section{Practical algorithm to perform an 
independent-bin Bayesian unfolding}
The basic trick to avoid the mentioned difficulty is to apply
Bayes' theorem to causes and effects, instead than to 
true and observed spectrum. In practice, instead of using
of Eq.~(\ref{eq:bayes1}) we start from
\begin{eqnarray}
P(C_i\,|E_j,\,I) &=& \frac{P(E_j\,|\,C_i,\,I)\cdot P(C_i\,|\,I)}
                         {\sum_i P(E_j\,|\,C_i,\,I)\cdot P(C_i\,|\,I)}\,,
\label{eq:bayes_CE} 
\end{eqnarray}
or
\begin{eqnarray}
\theta_{ij}&=& \frac{\lambda_{ji}\cdot P(C_i\,|\,I)}
         {\sum_i \lambda_{ji}\cdot P(C_i\,|\,I)}\,, 
\label{eq:bayes_CE_theta}
\end{eqnarray}
having defined $\theta_{ij} \equiv P(C_i\,|E_j,\,I)$
in analogy to $\lambda_{ji} \equiv P(E_j\,|\,C_i,\,I)$\,.

At this point a very important remark is in order. 
The prior  in Eqs.~(\ref{eq:bayes_CE})-(\ref{eq:bayes_CE_theta}) 
has a different meaning from that of Eq.~(\ref{eq:bayes1}). 
In Eq.~(\ref{eq:bayes1}) $P(\mvec x_C\,|\,I)$ assigns different probabilities
to all possible spectra.  Instead,
in  Eqs.~(\ref{eq:bayes_CE})-(\ref{eq:bayes_CE_theta}) 
  $P(C_i\,|\,I)$ stands for 
{\it a single spectrum} (more precisely, all spectra that 
differ from each other just by normalization). This can be better
understood analyzing the meaning of `uniform' (or `flat') 
referred to  $P(\mvec x_C\,|\,I)$ and to $P(C_i\,|\,I)$.
\begin{itemize}
\item 
 $P(\mvec x_C\,|\,I) = \mbox{\it constant}$ means 
{\it all spectra} are equally likely.
\item
 $P(C_i\,|\,I) = \mbox{\it constant}$ means we consider 
the cause bins equally
likely, i.e. the prior assess an initial belief in {\it flat spectra}. 
\end{itemize}
In other words, while a flat $P(\mvec x_C\,|\,I)$
means indifference about all possible spectra in order 
to `let the data speak by themselves', 
a flat $P(C_i\,|\,I)$ is a strong assumption that usually does 
not correspond to our priors concerning the physics case.
This implies that we have to tune somehow the algorithm in order
to take into account of this gross approximation. 
We will come back to this issue in Sec.~\ref{sec:iter}.

\begin{figure}[t]
\begin{center}
\begin{tabular}{c}
\epsfig{file=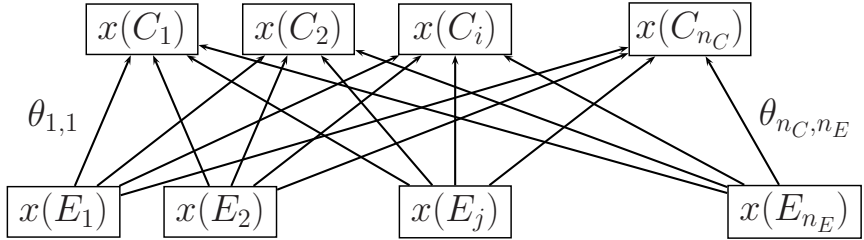,clip=,width=0.9\linewidth}
\end{tabular}
\end{center}
\caption{{\sl Sharing counts observed in effect-cells
among cause-cells according to $\theta_{ij} = P(C_i\,|\,E_j,\,I)$.}}
\label{fig:bn2}
\end{figure}

At this point, having evaluated  $P(C_i\,|\,E_j,\,I)$, 
we can use it to share the counts observed
in each effect-bin among  all cause-bins
(see Fig.~\ref{fig:bn2}).
 The estimate of the true spectrum 
is obtained repeating this sharing for all observed bins
and taking into account inefficiency. An uncertainty on the
unfolded spectrum is also evaluated. Let us see how all this
was done in the old algorithm and how it has been improved.

\subsection{Old algorithm (Ref.~\cite{BU})}
In the old algorithm the share of observed events among the causes
was done only considering
expectations and applying an inefficiency correction,
going through the following steps:
\begin{itemize}
\item
number of counts in $C_i$ due to the observation in $E_j$: 
\begin{eqnarray}
\left.x(C_i)\right|_{x(E_j)} & \approx & P(C_i\,|E_j,\,I) \cdot x(E_j) = 
                    \theta_{ij}\cdot x(E_j)\,;  \hspace{1.0cm}
\end{eqnarray}
\item
number of counts in $C_i$ due to all observations: 
\begin{eqnarray}
\left.x(C_i)\right|_{\mvec x_E} & \approx & 
               \sum_{j=1}^{n_E} P(C_i\,|E_j,\,I) \cdot x(E_j) = 
               \sum_{j=1}^{n_E}\theta_{ij}\cdot x(E_j)
\,; \hspace{0.5cm}
\end{eqnarray}
\item
number of counts in $C_i$ that also takes into account efficiency
($\epsilon_i$):
\begin{eqnarray}
x(C_i) & \approx &\frac{1}{\epsilon_i}\, \left.x(C_i)\right|_{\mvec x_E} 
 =  \frac{1}{\epsilon_i}\, 
               \sum_{j=1}^{n_E} P(C_i\,|E_j,\,I) \cdot x(E_j) 
\label{eq:old_estimate}
\\
&& \hspace{2.2cm}= 
       \frac{1}{\epsilon_i} \sum_{j=1}^{n_E}\theta_{ij}\cdot x(E_j)
\end{eqnarray}
where
\begin{eqnarray}
\epsilon_i &=& \sum_{j=1}^{n_E} P(E_j\,|\,C_i,\,I) = 
              \sum_{j=1}^{n_E} \lambda_{ji} =
              1 - \lambda_{n_E+1,\,i} \,. \label{eq:def_eps}
\end{eqnarray}
\item
Finally, we have to remember that  the several $\lambda_{ij}$ were
estimated by MC simulation as 
\begin{eqnarray}
\lambda_{ji}\approx x(E_j)^{MC}/x(C_i)^{MC}\,,
\end{eqnarray}
from which $\epsilon_i $ and $\theta_{ij}$ are calculated according
to Eqs. (\ref{eq:def_eps}) and (\ref{eq:bayes_CE_theta}).
\end{itemize}
Essentially, $x(C_i)$ of Eq.~(\ref{eq:old_estimate}) was considered 
an `estimator', whose `error' was calculated by standard
`error propagation' (this was a pragmatic compromise between `standard
methods' I was accustomed at that time 
and the Bayesian approach I was in the process of learning
 -- anyhow, not bad to begin, and not 
worst than other methods.)

At this point, there is still open the issue of the inappropriate priors,
that we have encountered  
at the beginning of this section. 
Since this question remains in the improved algorithm, 
it will be discussed later in Sec.~\ref{sec:iter}.

\subsection{Improvements}\label{ss:improvements}
As it is easy to understand from the previous
description, weak points of the 
old algorithm are the treatment of small numbers and 
the calculation of uncertainty by `error propagation
formulas'. 
Let us see how we can improve them. 
\begin{itemize}
\item 
Instead of just `estimate' $\lambda_{ij}$ as  
$x(E_j)^{MC}/x(C_i)^{MC}$, we can model their 
pdf by a Dirichlet (see Appendix A): 
\begin{eqnarray}  
\mvec \lambda_i & \sim & \mbox{Dir}[\mvec \alpha_{prior}\, + \, 
                      \left.\mvec x_E^{MC}\right|_{x(C_i)^{MC}}]\,,
\label{eq:lambda_i_dir}
\end{eqnarray}
where 
\begin{itemize}
\item 
$\mvec \alpha_{prior}$ is the initial set of Dirichlet 
parameters -- typically unitary for uniform priors 
 (see Appendix B for details); 
\item
$\left.\mvec x_E^{MC}\right|_{x(C_i)^{MC}}$
stands for the numbers of MC counts generated in $C_i$ 
and which end in each of 
{\it all} possible $n_E+1$ `effects' 
(i.e. we also have to consider the inefficiency bin --
see Fig.~\ref{fig:unfnet} -- in order to satisfy the condition 
$\sum_i\lambda_i = 1$
 of the Dirichlet variables).
\end{itemize}
\item
The uncertainty about $\Lambda$ is taken into account by {\it sampling}
its values, that is equivalent to perform the integral 
(\ref{eq:marg_lambda}).\footnote{Note that if we are in doubt
about several smearing matrices, because e.g. they are obtained
from different parameters of the simulation, the 
(`systematic') effect of this uncertainty can be taken
into account sampling from the different $\Lambda$'s, with 
weights depending on our confidence on each of them.} 
Therefore, for  each sampling, 
   \begin{itemize}
   \item the smearing matrix elements $\lambda_{ji}$ are extracted  
         according to a Dirichlet distribution;
   \item the efficiencies $\epsilon_i = \sum_{j=1}^{n_E}\lambda_{ji}$ 
         are calculated;
   \item the inverse probabilities  $\theta_{ij}$ are obtained 
         from the Bayes formula. 
   \end{itemize}
\item 
The sharing of the number of counts $x(E_j)$ observed in an effect-cell
among all cause-bins is performed using a 
multinomial distribution (see Appendix A.1):
\begin{eqnarray}  
\left.\mvec x_C\right|_{x(E_j)} & \sim & 
\mbox{Mult}[x(E_j),\,\mvec \theta_j]\,,
\label{eq:extract_xC}
\end{eqnarray}
with  $\mvec \theta_j = \{\theta_{1,j},\,
\theta_{2,j},\,\ldots,\,\theta_{n_C,j},\,\}$.
\item
The  contributions of all {\it observed} 
effect-bins are summed up, i.e. 
 \begin{eqnarray}  
\left.\mvec x_C\right|_{\mvec x_E} 
& = & \sum_{j=1}^{n_E} \left.\mvec x_C\right|_{x(E_j)}\,.
\label{eq:x_C_obs}
\end{eqnarray}
Inefficiency are taken into account as in the old algorithm:
   \begin{eqnarray}  
   \mvec x_C
   & = & \frac{\left.\mvec x_C\right|_{\mvec x_E}}
              {\mvec \epsilon }\,,
   \label{eq:x_C_mu_eps}
   \end{eqnarray}
   where the ratio in  the r.h.s is done component by component.  
\end{itemize}
However, we need to remember that the observed number of events
in each effect bin, $x(E_j)$,
 comes from a Poisson distribution with unknown 
parameter $\mu_j$, whose inference can be conveniently performed
starting from a  conjugate gamma distribution (see e.g. Ref.~\cite{BR}).
It follows that 
\begin{eqnarray}
\mu_j & \sim & \mbox{Gamma}[c_j+x(E_j),\, r_j+1]\,,
\label{eq:mu_j_Gamma}
\end{eqnarray}
where $c_j$ and $r_j$ are the initial parameters of the gamma. 
[The case of a flat prior corresponds to 
$c_j=1$ and $r_j\rightarrow 0$, i.e. an exponential with 
vanishing rate.] 

Therefore, what should be shared among 
the cause-bins is not $x(E_j)$ but  
$\mu_j$,  extracted in each sampling 
according to Eq.~(\ref{eq:mu_j_Gamma}). 
However, $\mu_j$ is a continuous variable and therefore
cannot be used  
as `trials parameter' of a multinomial distribution.
Nevertheless,
fractional events can indeed be shared making use of a little trick:
   \begin{itemize}
   \item  
   $\mu_j$ is extracted according to Eq.~(\ref{eq:mu_j_Gamma}) and 
   it is rounded to the closest {\it positive} integer, 
   that we indicate here by $m_j$: 
   $m_j$ will be the number of trials used in the multinomial random
   generator.
   \item 
   $\left.\mvec x_C\right|_{m_j}$ is extracted and then 
   rescaled by the factor $\mu_j/m_j$, i.e.
    \begin{eqnarray}
     \left.\mvec x_C\right|_{m_j}  & \sim & 
     \mbox{Mult}(m_j,\,\mvec \theta_j)\,, \\
    \left.\mvec x_C\right|_{\mu_j} &=& \frac{\mu_j}{m_j}\,
      \left.\mvec x_C\right|_{m_j}\,.
    \end{eqnarray}
\end{itemize}
This means that  Eq.~(\ref{eq:x_C_obs}) has to be replaced by
\begin{eqnarray}  
\left.\mvec x_C\right|_{\mvec x_E} 
& = & \sum_{j=1}^{n_E} \left.\mvec x_C\right|_{\mu_j}\,;
\label{eq:x_C_mu}
\end{eqnarray}
then the inefficiency corrections are applied as usual. 

Each sampling provides a spectrum $\mvec x_C^{(t)}$, where 
$t$ is the `time' index of the sampling. After $N$ samplings
we can calculate an average spectrum, variance 
and covariance for each cause-bin, as well as covariances
among bins, and any other statistical summary; or we can inspect 
graphically each $x(C_i)$.

This algorithm has been implemented in the R language\,\cite{R}
and it is freely available on the web~\cite{R-code}. 
The R language has been chosen
because it is a powerful 
scripting language,\footnote{I take the opportunity to make a point
about the teaching of scripting languages, and in particular of
$R$, in the physics courses. Surely physics students might need to learn 
C (and perhaps C++) at some point, but it is a matter of fact that writing 
some C code to solve little/medium problems that occur everyday,
and that might also need some graphics,
is a pain, not only for students. The consequence is that
students usually do not write little programs in C to solve
the problems they meet in the general physics  or 
laboratory courses, continuing to use, instead,  spread-sheets,
learned in  high school (and forget real programming until they
need it for their theses, if they ever will need it).
In my opinion, teaching a scripting language
like R, perhaps in parallel to C, would be a good solution.
(but if I really had
to chose, I would opt for R, as the language to begin).
(Personally, since I have discovered scripting languages,
I use C  only for though professional tasks. 
Indeed, I have even almost abandoned {\it Mathematica},
unless I really need symbolic calculation.)
}
open source, multi-platform, oriented towards statistics
applications, well 
maintained and with tons of contributed extension packages.
In practice a kind of programming {\it lingua franca}. 

In order to test the 
algorithm, also a simple event generator has been included, 
that reproduces the toy models of Ref.~\cite{BU}. 
Some results will be presented in Sec.~\ref{sec:results}

\section{Iterations and smoothing}\label{sec:iter}
Let us now go back to the issue of the priors, 
that we have left open from the beginning of the 
previous section. The crucial thing to understand 
 is that, as we stated above, 
{\it instead of using a prior flat over the possible spectra, 
we are using a particular, flat spectrum as prior}. Therefore, 
the posterior [i.e. the ensemble of $\mvec x_C^{(t)}$
obtained by sampling] is affected by this quite 
strong assumption, that seldom holds in real cases.

Obviously, the first idea a practical physicist has,
in order to fix the problem, is to do some {\it fine tuning}.
In fact, simulations on toy models show that
the unfolded distribution reproduces rather well 
the true one, even with a flat spectrum as prior.
However, it still `remembers the flat prior'. 
This effect can be cured iterating the procedure, using the posterior
as prior in a subsequent unfolding. 
Empirically we learn that, in `normal' cases, just 
two or three steps are sufficient to recover quite accurately
the true spectrum.

As it has been discussed in length in Ref.~\cite{BU},
we cannot repeat the iterations for a very long time,
otherwise there will be a kind of positive feedback, driven 
by fluctuations, that makes the asymptotically
unfolded spectrum `crazy' (it simply means that it is
far away from all reasonable expectations). Again, 
as we understand it, it is a question of 
judging the outcome by our physics priors,\footnote{Good 
physicists do have priors and always use them! 
(Only the perfect idiot has no priors.)}
that, although rather vague, tell us that, given  
the kind of measurement, a spectrum with wild oscillations  
is far from our beliefs. This kind of problem happens also with
other kind of unfolding algorithms
and it is usually cured by {\it regularization}.

Regularization is based on the subjective scientific prejudice 
that `wild' distributions are not physical
(and, as the reader knows or might imagine, this very reasonable 
subjective ingredient is unavoidably also
embedded in all non-Bayesian methods).
In practice, regularization can be implemented constraining
adjacent bins not to be `too discontinuous', for example limiting
the absolute value of the derivatives calculated numerically
on the unfolded spectrum.
Or one might fit the spectrum with some orthonormal functions, 
but dumping `high frequencies'. And so on. 

I do not have a strong opinion on the matter. My 
recommendation is that one should know what one is doing.
My preferred regularization method 
for one-dimensional problems consists 
in smoothing the posterior before injecting it 
as prior of the next iteration 
(but not after the last step!).
This smoothing can be performed by a low order polynomial fit,
as it is implemented in the demo scripts accompanying the program.  
I find this technique quite simple, well consistent with 
the spirit of this unfolding method and with the reasons 
to go through the iterations. 
Moreover, it is easy to understand that the 
procedure unfolding-smoothing-unfolding converges rapidly. 

In conclusion, 
although the idea of iteration does not seem very
Bayesian (we cannot squeeze the same data twice!),
it is in fact just an adaptive way to recover what 
could have been obtained by a more reasonable uniform 
prior over the possible spectra. (Stated with other words,
it would be non-Bayesian to stick to the 
first iteration, obtained from a prior that rarely
corresponds to physics spectra!) For this reason the 
idea of iteration has been kept, although some effort has 
been done to get rid of it. 
In fact, in the new code the user can,
instead of providing a (typically flat) spectrum as prior, 
set up a function [\,{\tt priorf()}\,] 
that returns a normalized spectrum 
chosen at random according to a probability distribution
describing prior knowledge.
In this way, in each sampling also the prior is sampled. 
However, I anticipate that implementing the function  {\tt priorf()}
might not be that trivial, and iterative procedure still
remains a pragmatic compromise 
to slalom among the difficulties. 

\section{Results on toy models}\label{sec:results}
\begin{figure}
\begin{center}
\epsfig{file=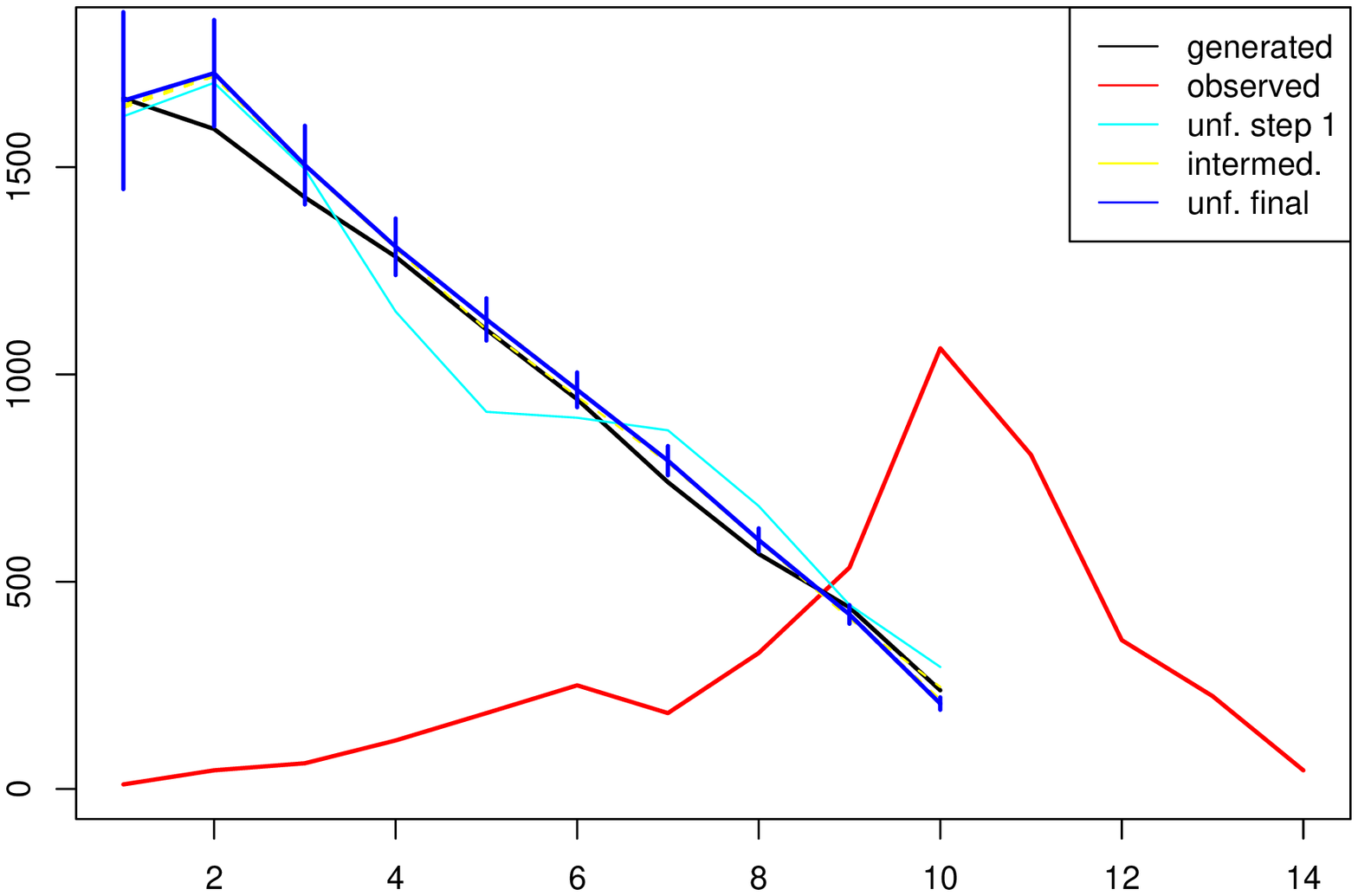,bb=40 240 555 590,clip=,width=0.47\linewidth} 
\hspace{0.5cm}
\epsfig{file=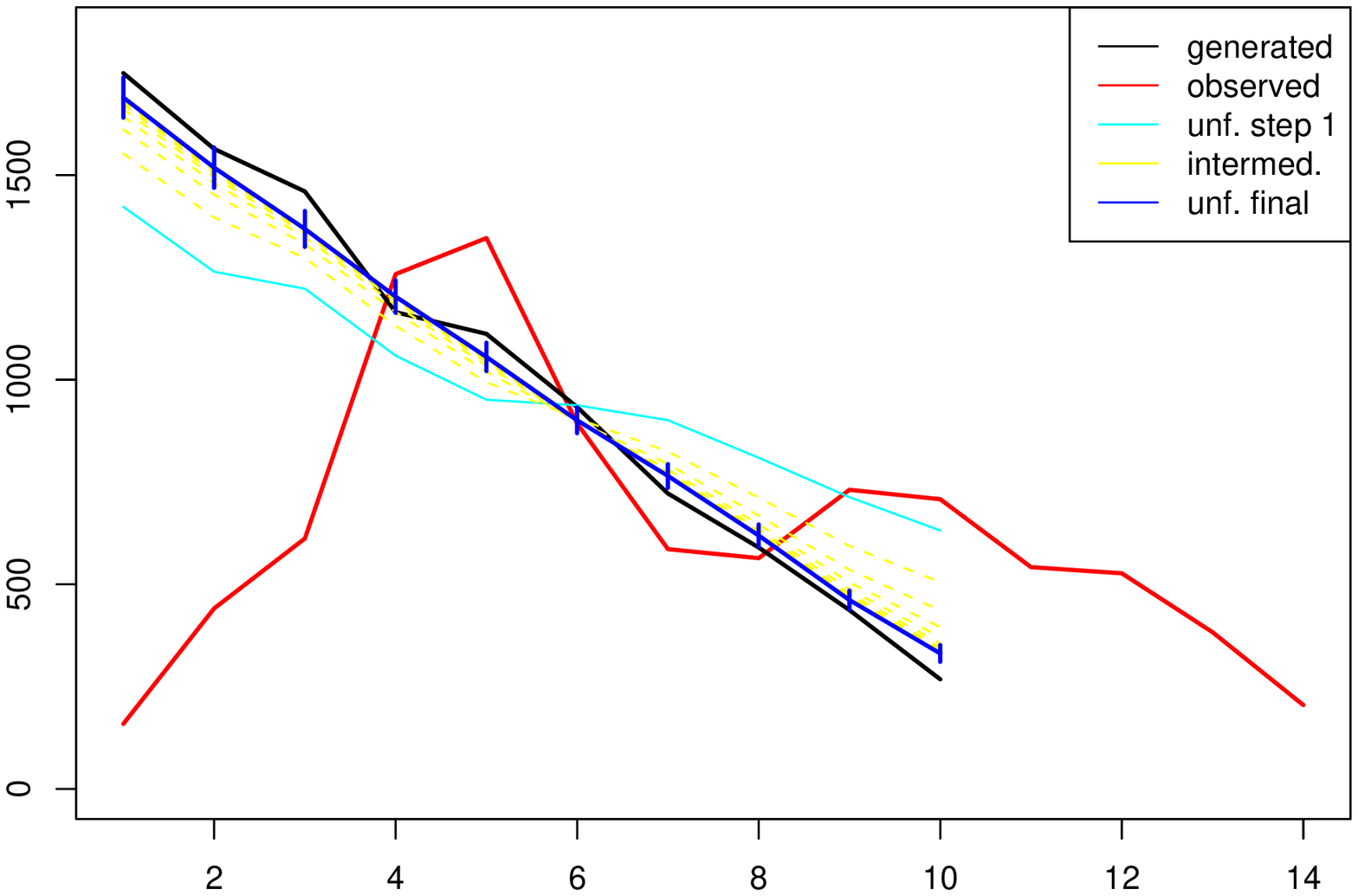,bb=40 240 555 590,clip=,width=0.47\linewidth} \\
\epsfig{file=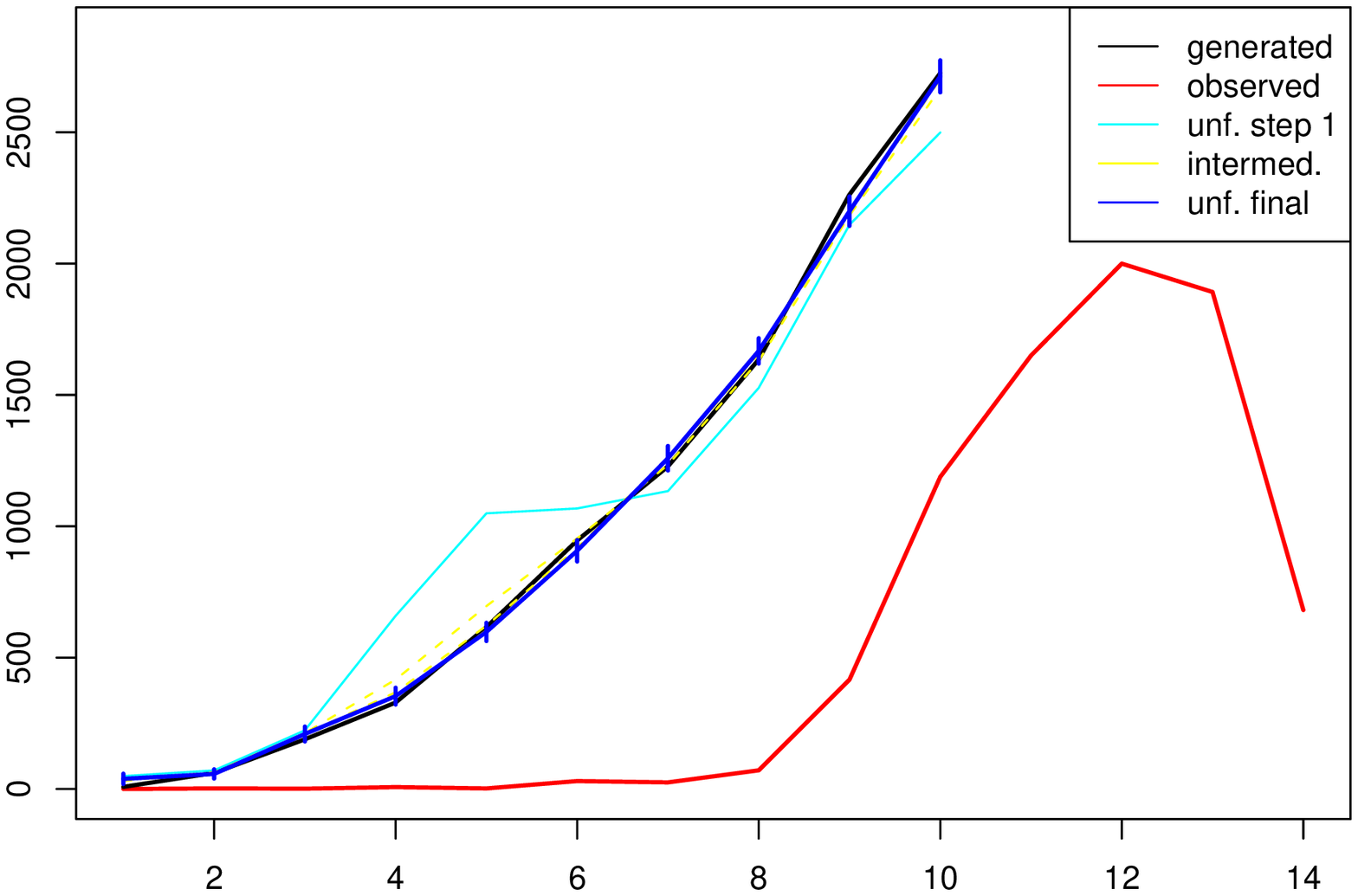,bb=40 240 555 590,clip=,width=0.47\linewidth} 
\hspace{0.5cm}
\epsfig{file=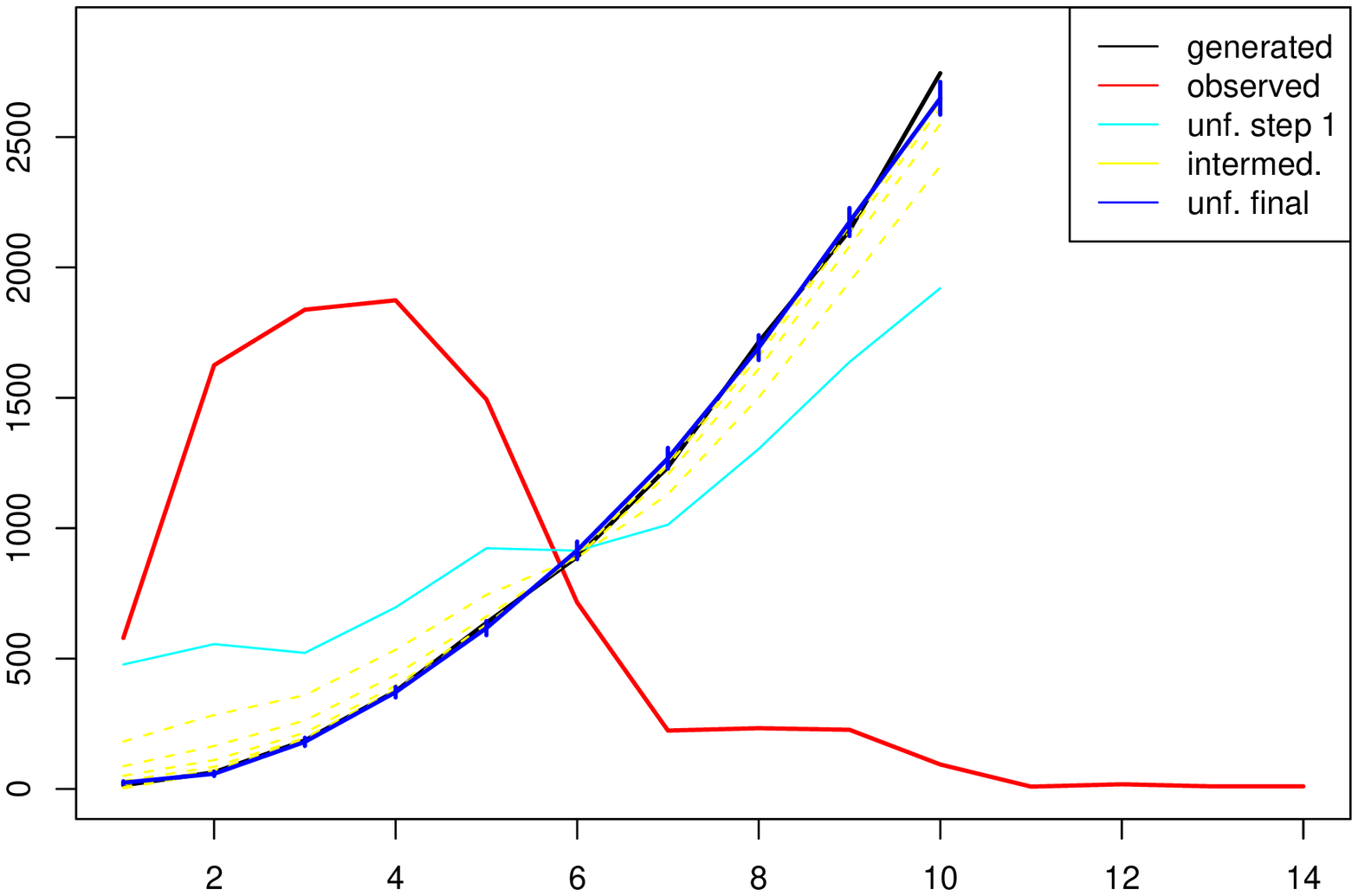,bb=40 240 555 590,clip=,width=0.47\linewidth} \\
\epsfig{file=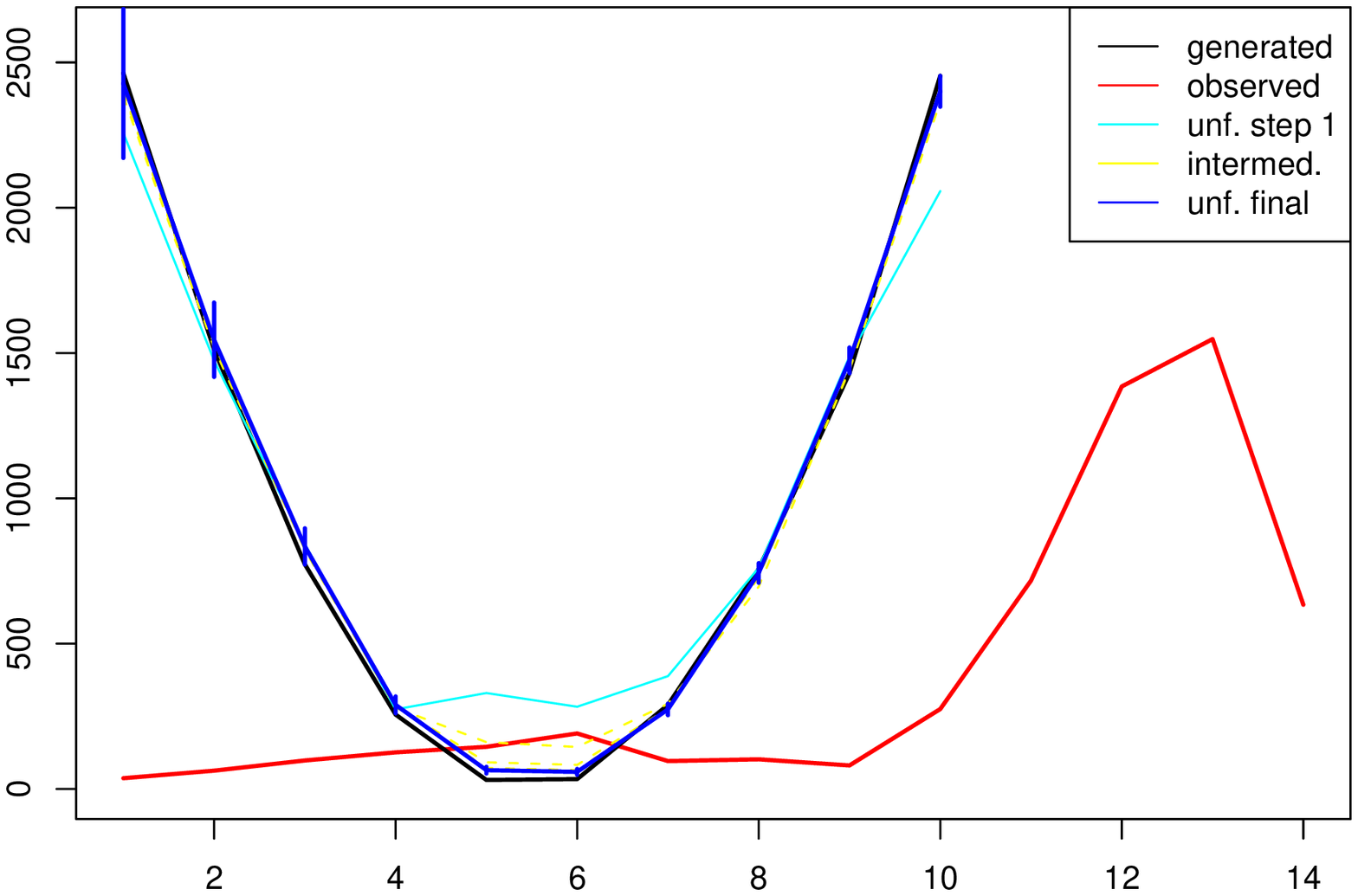,bb=40 240 555 590,clip=,width=0.47\linewidth} 
\hspace{0.5cm}
\epsfig{file=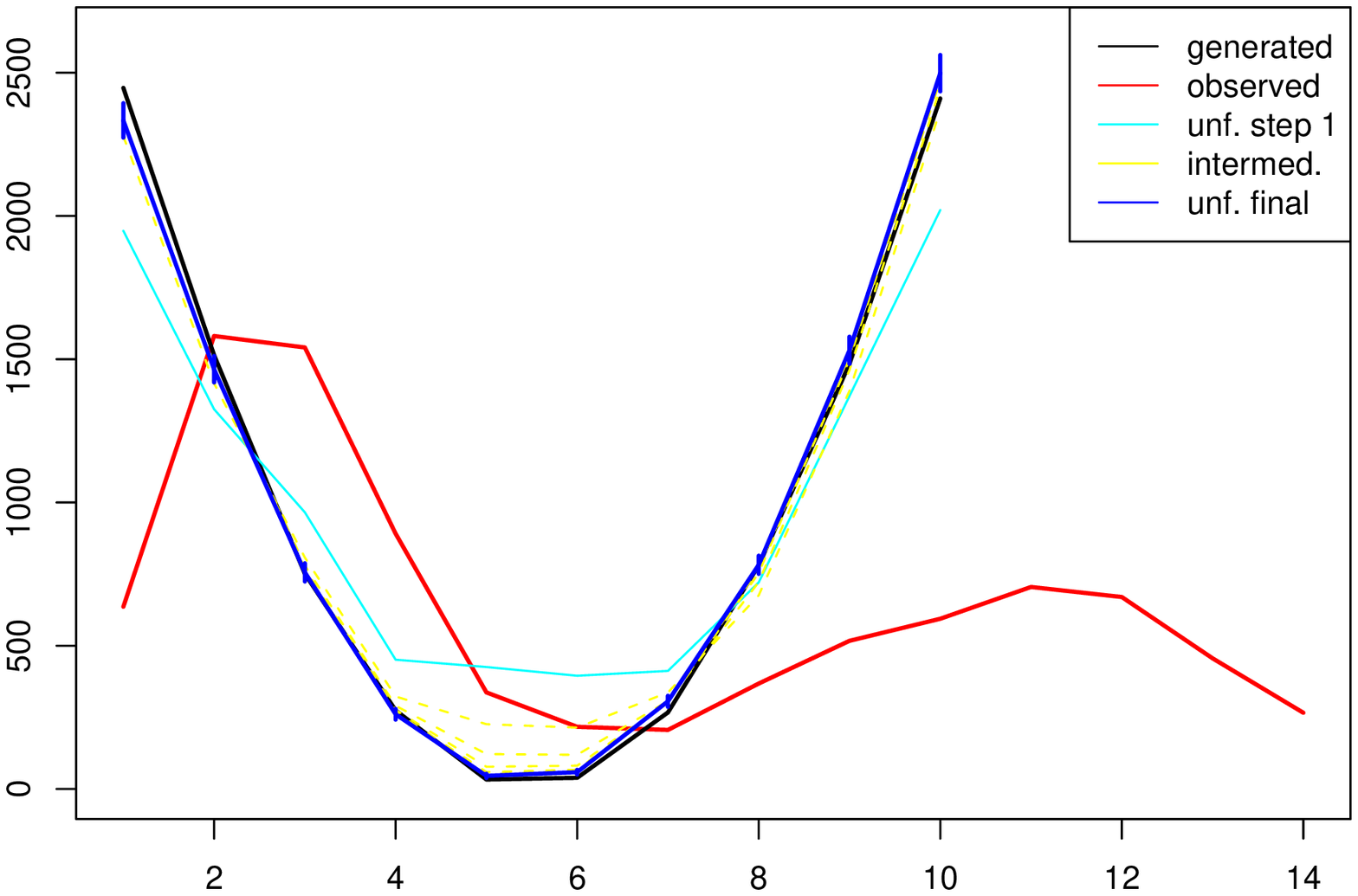,bb=40 240 555 590,clip=,width=0.47\linewidth} \\
\epsfig{file=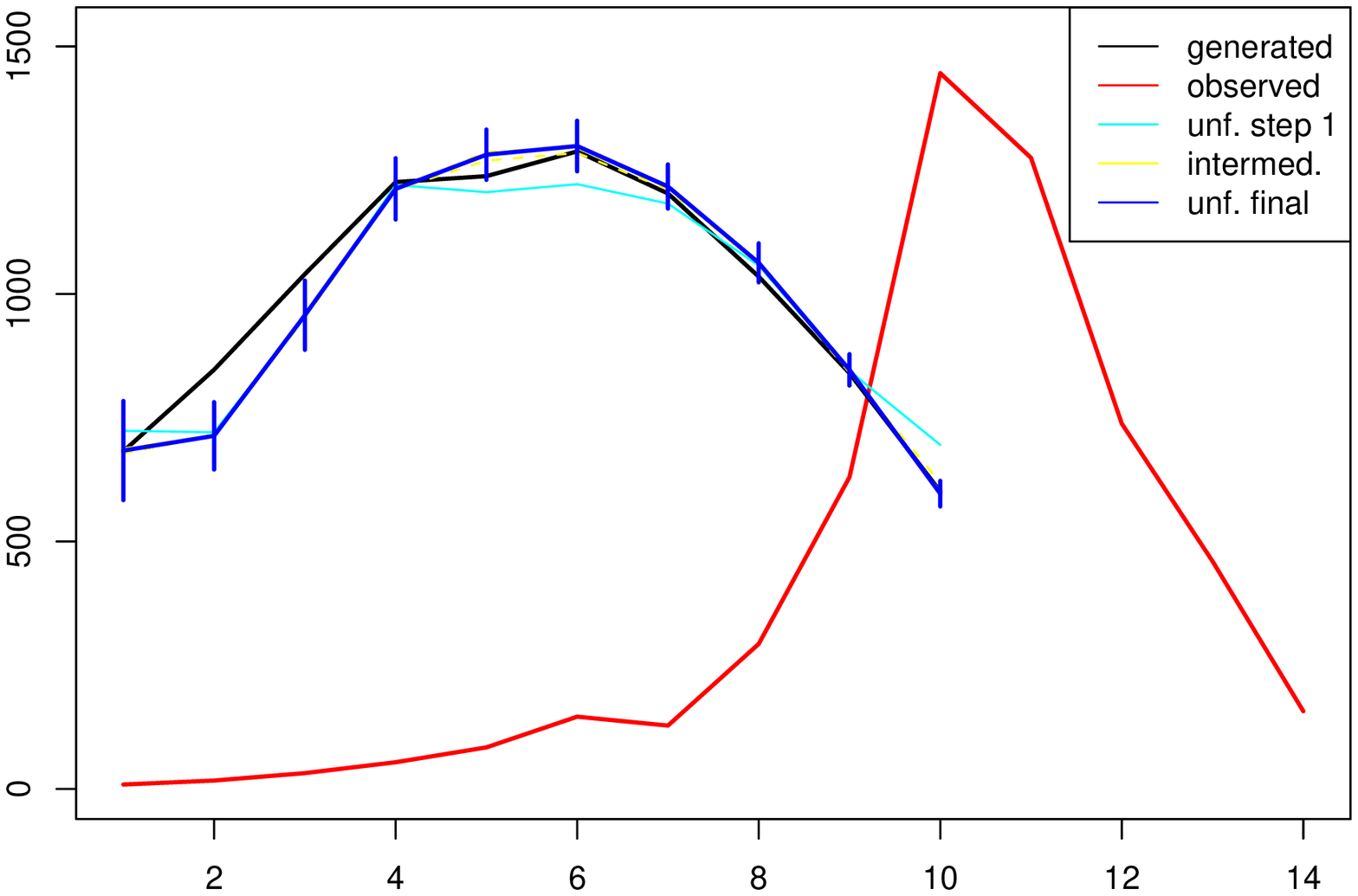,bb=40 240 555 590,clip=,width=0.47\linewidth} 
\hspace{0.5cm}
\epsfig{file=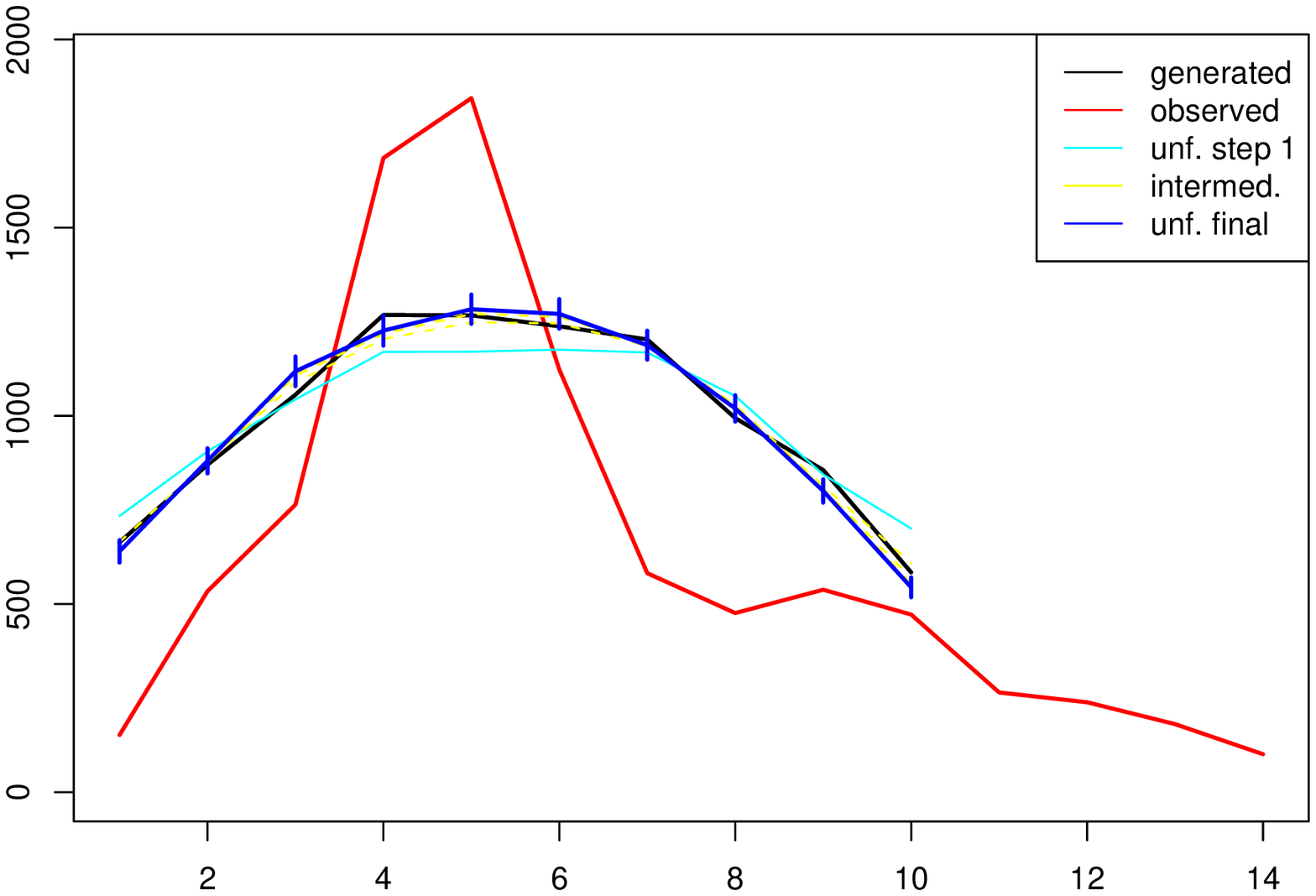,bb=40 240 555 590,clip=,width=0.47\linewidth} 
\end{center}
\caption{\sl Result of unfolding applied to four toy models 
({\rm data.func} parameter from 1 to 4 in the demo program) and two different 
smearing matrices (left and right figures correspond, respectively,
to `Smearing 1' and `Smearing 2' of Tab.~\ref{tab:smearing}.).
}
\label{fig:toy1}
\end{figure}

The program has been checked with the same toy smearing 
matrices used in Ref.~\cite{BU}, reproduced here in 
Tab.~\ref{tab:smearing}. 
\begin{table}
{\small 
\begin{center}
\begin{tabular}{|c|cccccccccc|}
\multicolumn{11}{c}{Smearing 1} \\
\multicolumn{11}{c}{} \\ 
\hline
& $C_1$& $C_2$&  $C_3$& $C_4$&  $C_5$&  $C_6$& $C_7$&  $C_8$& $C_9$&  $C_{10}$  \\  
\hline
$E_1$& 0.010& 0& 0& 0& 0& 0& 0& 0& 0& 0 \\
$E_2$& 0.025& 0& 0& 0& 0& 0& 0& 0& 0& 0 \\
$E_3$& 0.040& 0& 0& 0& 0& 0& 0& 0& 0&  0 \\
$E_4$& 0.025& 0.050& 0& 0& 0& 0& 0&  0& 0&  0 \\
$E_5$& 0& 0.100& 0& 0& 0& 0& 0&  0& 0&  0 \\
$E_6$& 0& 0.050& 0.120& 0& 0& 0& 0& 0& 0& 0 \\
$E_7$& 0& 0& 0.120& 0& 0& 0& 0& 0& 0 & 0 \\
$E_8$&  0& 0& 0.060& 0.160& 0.025& 0& 0& 0& 0&  0 \\
$E_9$&  0& 0& 0& 0.160& 0.100& 0.120& 0.140& 0& 0&  0\\
$E_{10}$ &  0& 0& 0& 0.080& 0.250& 0.360& 0.280&  0.200& 0& 0 \\
$E_{11}$ &  0& 0& 0& 0& 0.125& 0.120& 0.280&  0.400& 0.225&  0 \\
$E_{12}$ &  0& 0& 0& 0& 0& 0& 0&  0.200& 0.450& 0.250 \\
$E_{13}$ &  0& 0& 0& 0& 0& 0& 0&  0& 0.225&  0.500 \\
$E_{14}$ &  0& 0& 0& 0& 0& 0& 0&  0& 0&  0.250 \\
\hline
$\epsilon$ & {\it 0.1} & {\it 0.2} & {\it 0.3}& {\it 0.4} & {\it 0.5} &
                  {\it 0.6} & {\it 0.7} & {\it 0.8}& {\it 0.9} & {\it 1}\\ 
\hline
\multicolumn{11}{c}{} \\ 
\multicolumn{11}{c}{} \\ 
\multicolumn{11}{c}{} \\ 
\multicolumn{11}{c}{Smearing 2} \\
\multicolumn{11}{c}{} \\ 
\hline
& $C_1$& $C_2$&  $C_3$& $C_4$&  $C_5$&  $C_6$& $C_7$&  $C_8$& $C_9$&  $C_{10}$  \\  
\hline
$E_1$& 0.045& 0 & 0 & 0 & 0 & 0 & 0 & 0 &  0 & 0.225 \\
$E_2$& 0.090& 0.054 & 0 & 0 & 0 & 0 & 0 & 0 & 0.225 & 0.405  \\
$E_3$& 0.045& 0.072 &  0.045 & 0 & 0 & 0 & 0 &  0.225& 0.360& 0.225 \\
$E_4$& 0 & 0.054& 0.045& 0.045& 0.225& 0.180& 0.360& 0.405 &0.225 & 0 \\
$E_5$&  0& 0& 0& 0.180& 0.360& 0.450& 0.315& 0.225& 0& 0 \\
$E_6$& 0& 0& 0& 0.315& 0.180& 0.180 &0.180 &0 &0.045 &0 \\
$E_7$&  0& 0& 0.135& 0.270& 0.045& 0& 0& 0.045& 0& 0 \\
$E_8$&  0& 0.045& 0.270& 0.045& 0& 0& 0.045& 0& 0.045& 0 \\
$E_9$& 0& 0.135& 0.270& 0& 0.045& 0.045& 0& 0& 0 &0.045 \\
$E_{10}$& 0& 0.360& 0.045& 0& 0.045& 0.045& 0&0& 0& 0 \\
$E_{11}$ & 0.180& 0.180& 0& 0& 0& 0& 0& 0& 0& 0 \\
$E_{12}$ & 0.270& 0& 0& 0.045& 0& 0& 0& 0& 0& 0 \\
$E_{13}$ & 0.270& 0& 0.045& 0& 0& 0& 0& 0& 0& 0 \\
$E_{14}$ & 0.090& 0& 0.045& 0& 0& 0& 0& 0& 0& 0 \\
\hline
$\epsilon$ & {\it 0.9} & {\it 0.9} & {\it 0.9}& {\it 0.9} & {\it 0.9} & 
                  {\it 0.9} & {\it 0.9} & {\it 0.9}& {\it 0.9} & {\it 0.9}\\ 
\hline
\end{tabular}
\end{center}
}
\caption{{\sl Smearing matrices $P(E_j\,|\,C_i)$ of the toy models}}
\label{tab:smearing}
\end{table}
The results of the simulations are given in Fig.~\ref{fig:toy1}. 
The generated 
distributions are shown in black.
The `measured' distributions (in red) 
look completely different from the `true', 
due to the very severe smearing matrices used. 
The figures also show 
the results after the first iteration
(light blue) and the intermediate ones 
(yellow). Note that, although in all simulations
twenty iterations have been performed, only a few intermediate
iterations are visible, because the others overlap with the
final step. This allows you to get a feeling 
about the speed of convergence of the algorithm, depending 
on the difficulty of the problem
 -- let us
remind that, thanks to the intermediate smoothing the algorithm
{\it does} converge.

\begin{figure}
\begin{center}
\epsfig{file=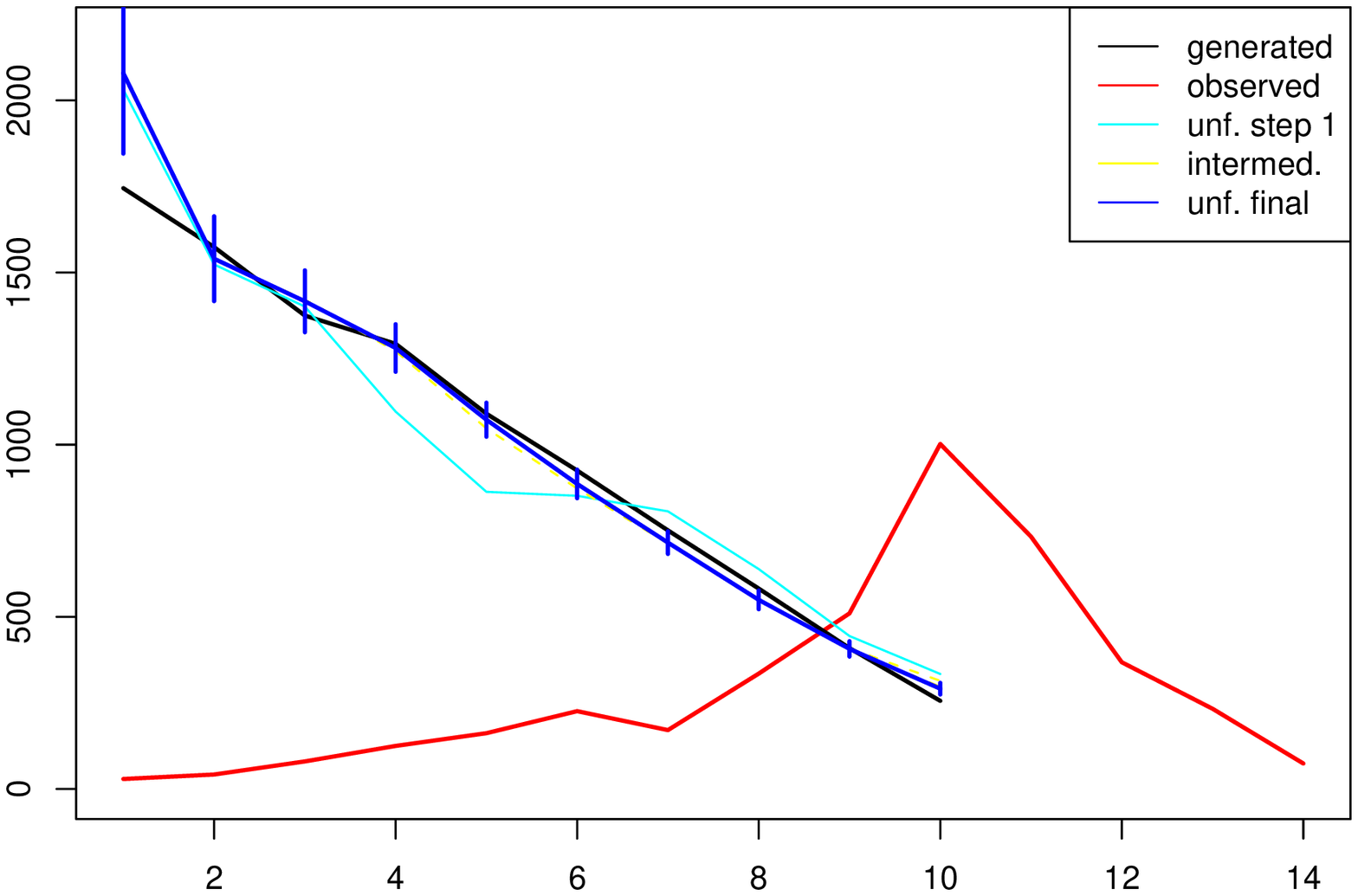,bb=40 240 555 590,clip=,width=0.47\linewidth} 
\hspace{0.5cm}
\epsfig{file=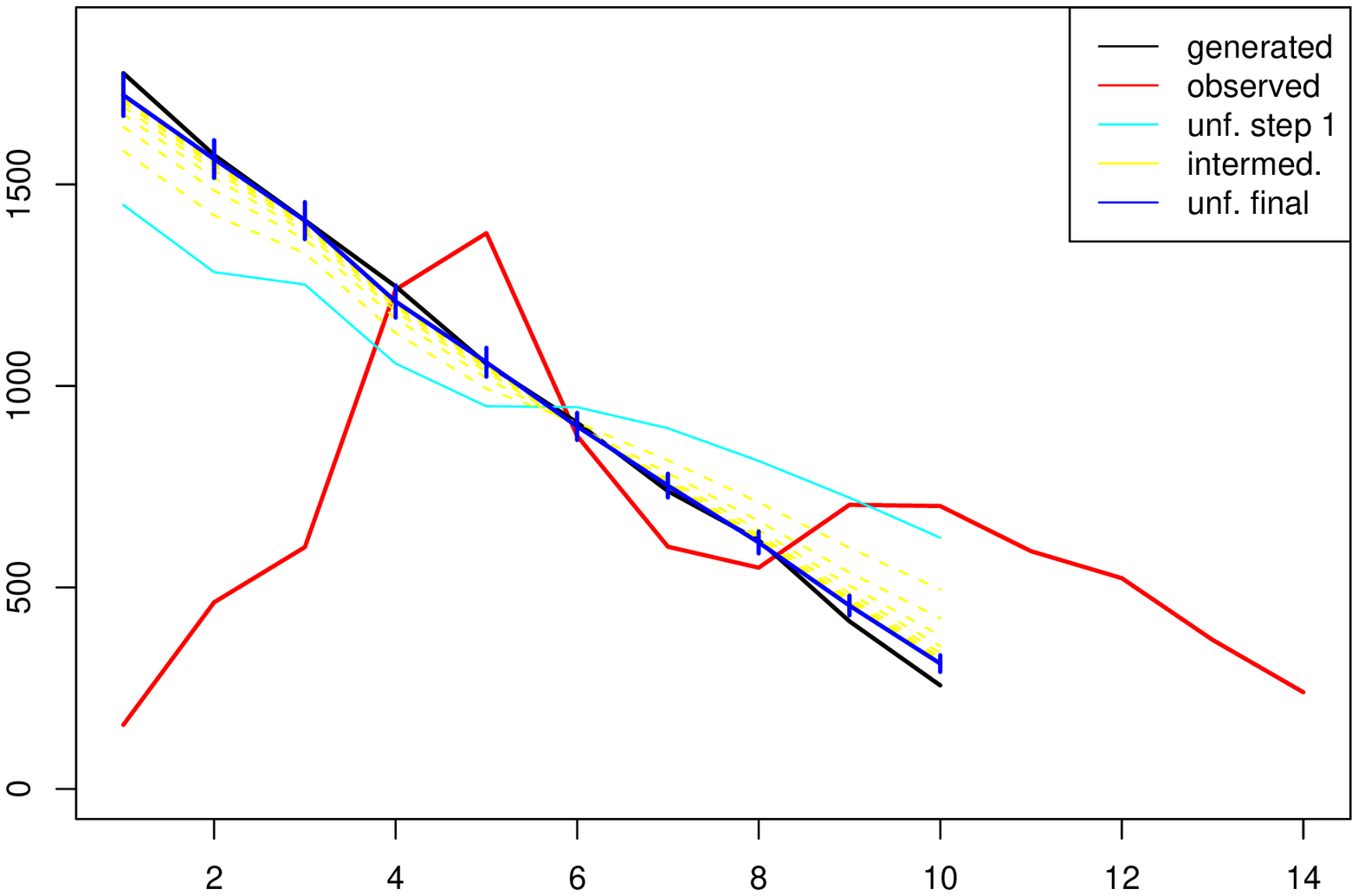,bb=40 240 555 590,clip=,width=0.47\linewidth} \\
\epsfig{file=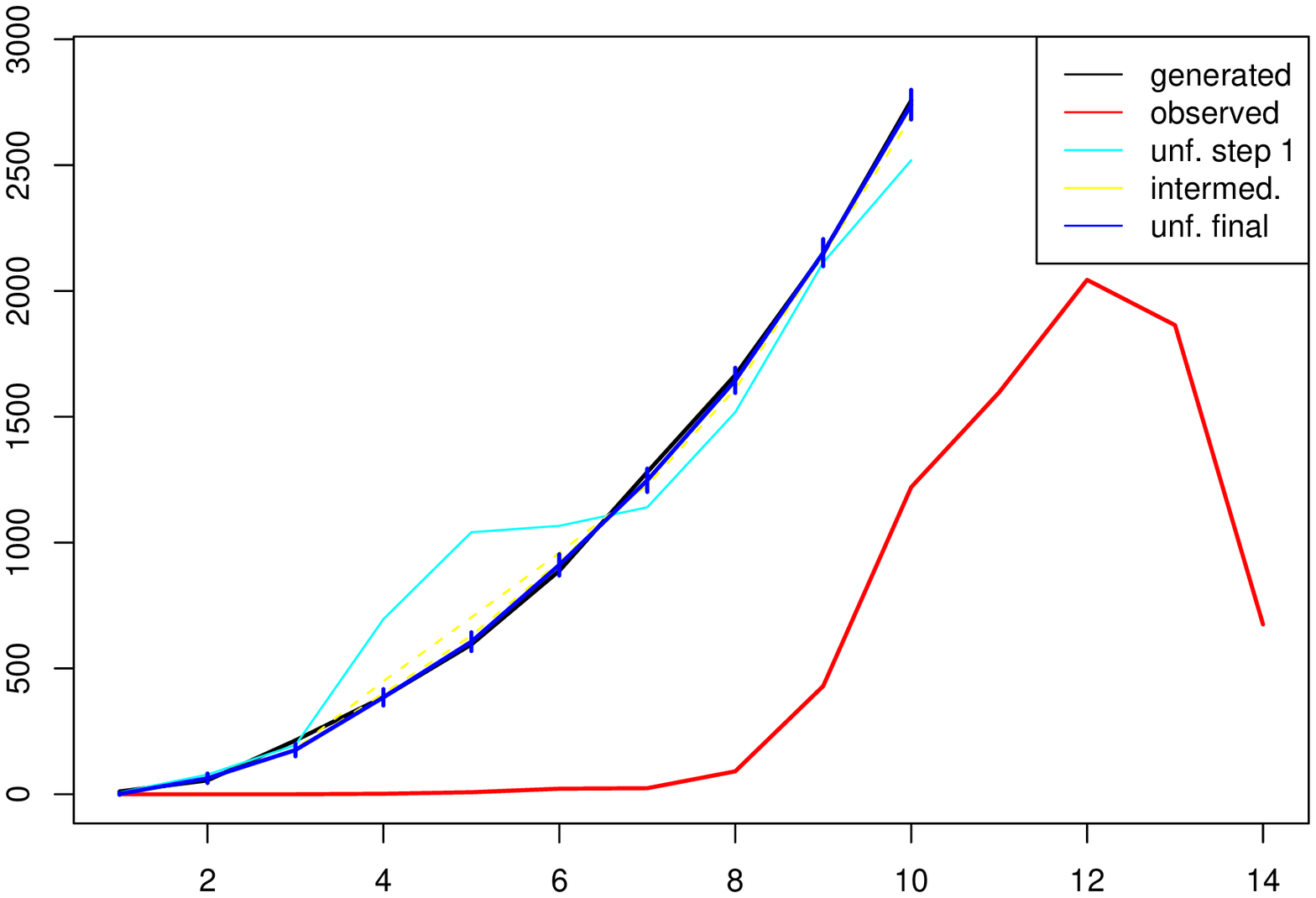,bb=40 240 555 590,clip=,width=0.47\linewidth} 
\hspace{0.5cm}
\epsfig{file=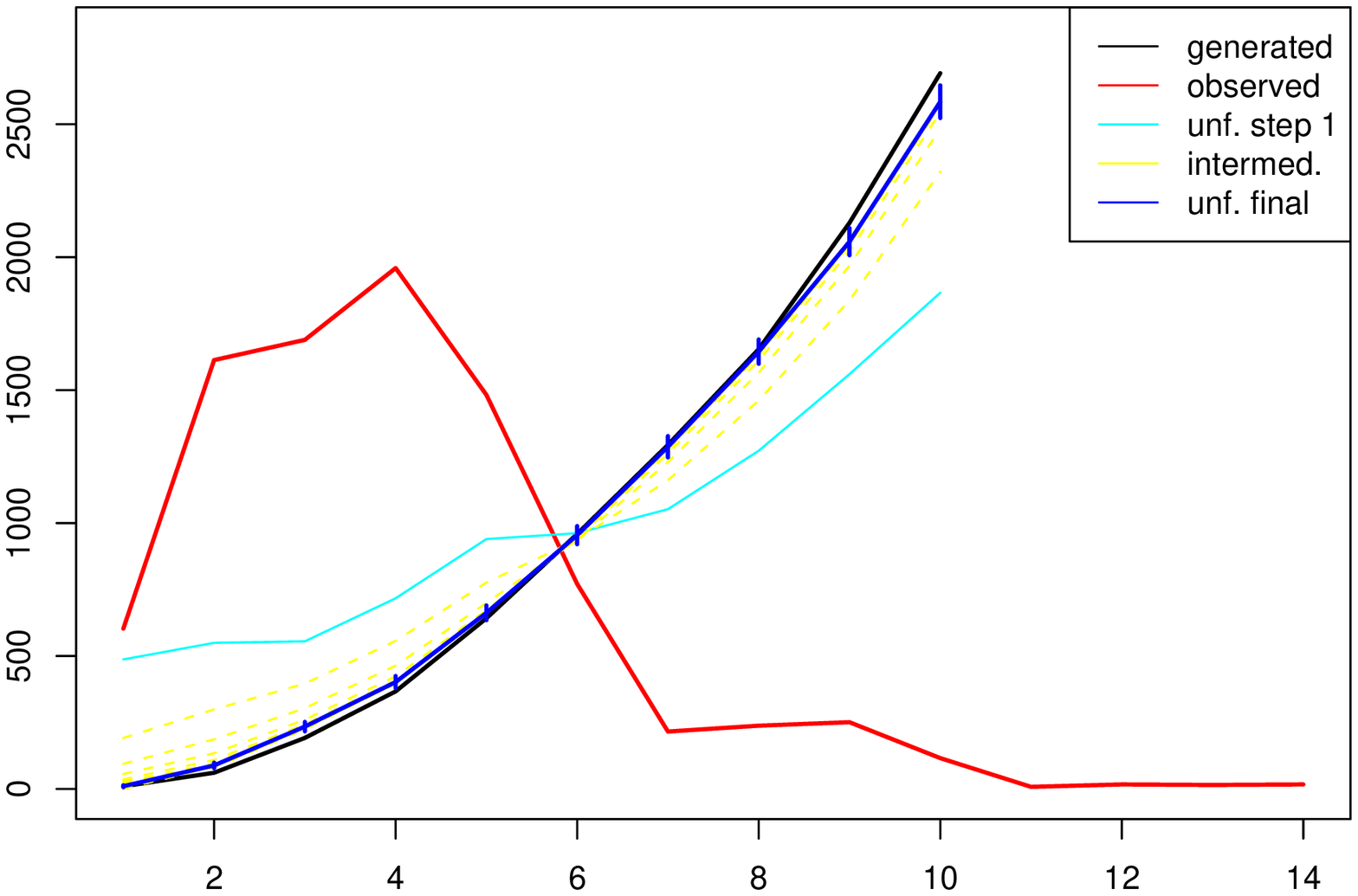,bb=40 240 555 590,clip=,width=0.47\linewidth} \\
\epsfig{file=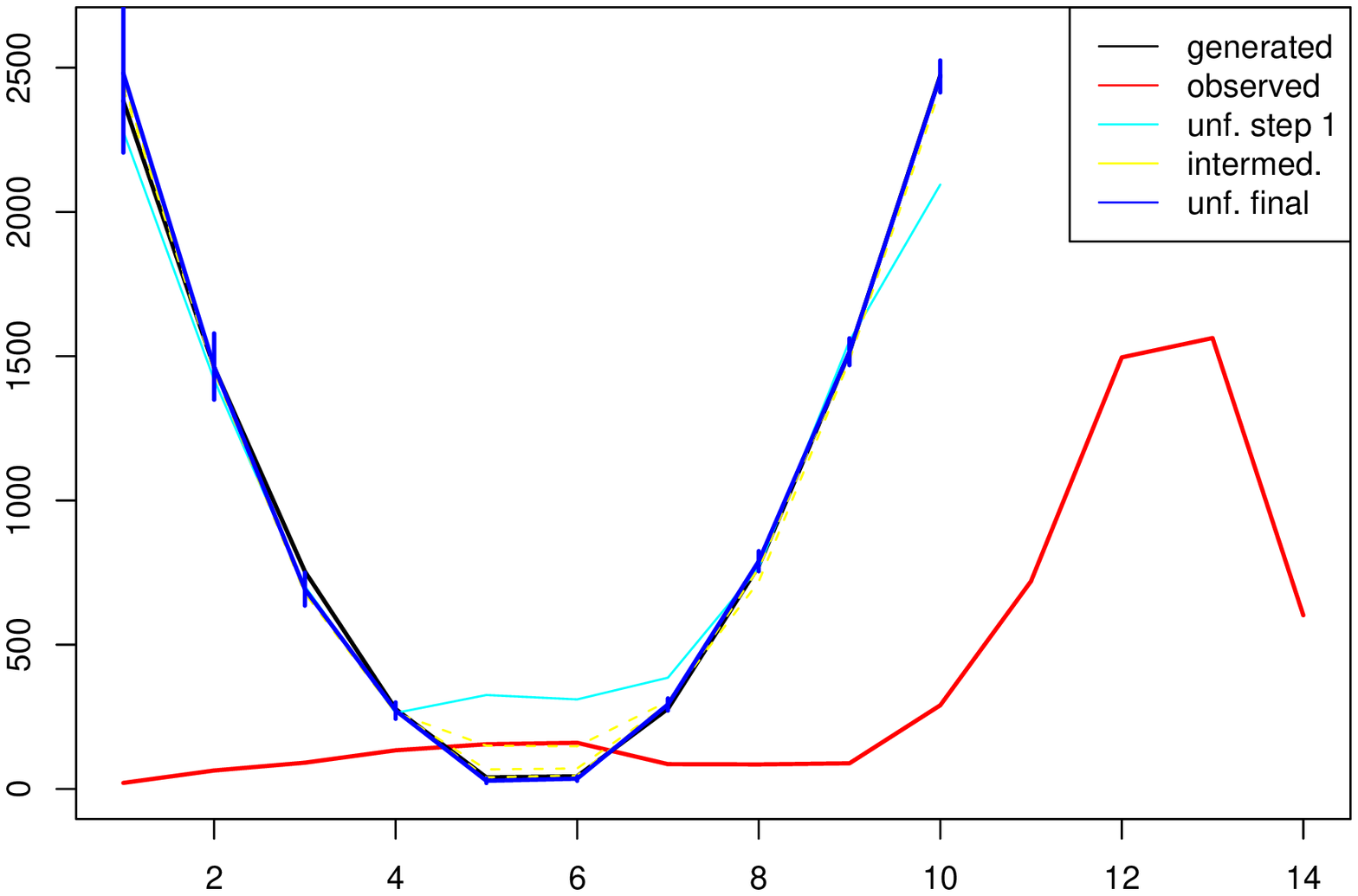,bb=40 240 555 590,clip=,width=0.47\linewidth} 
\hspace{0.5cm}
\epsfig{file=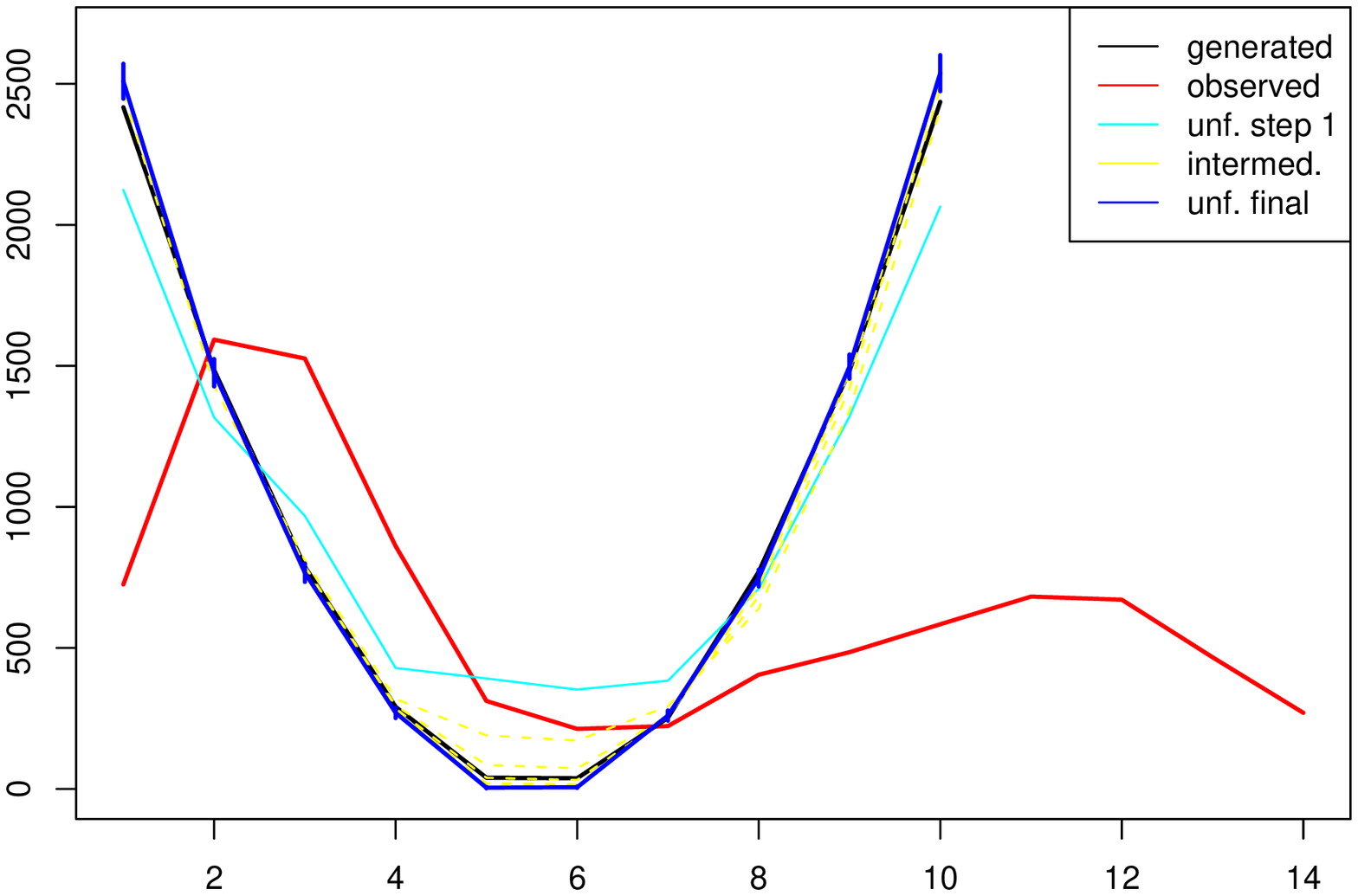,bb=40 240 555 590,clip=,width=0.47\linewidth} \\
\epsfig{file=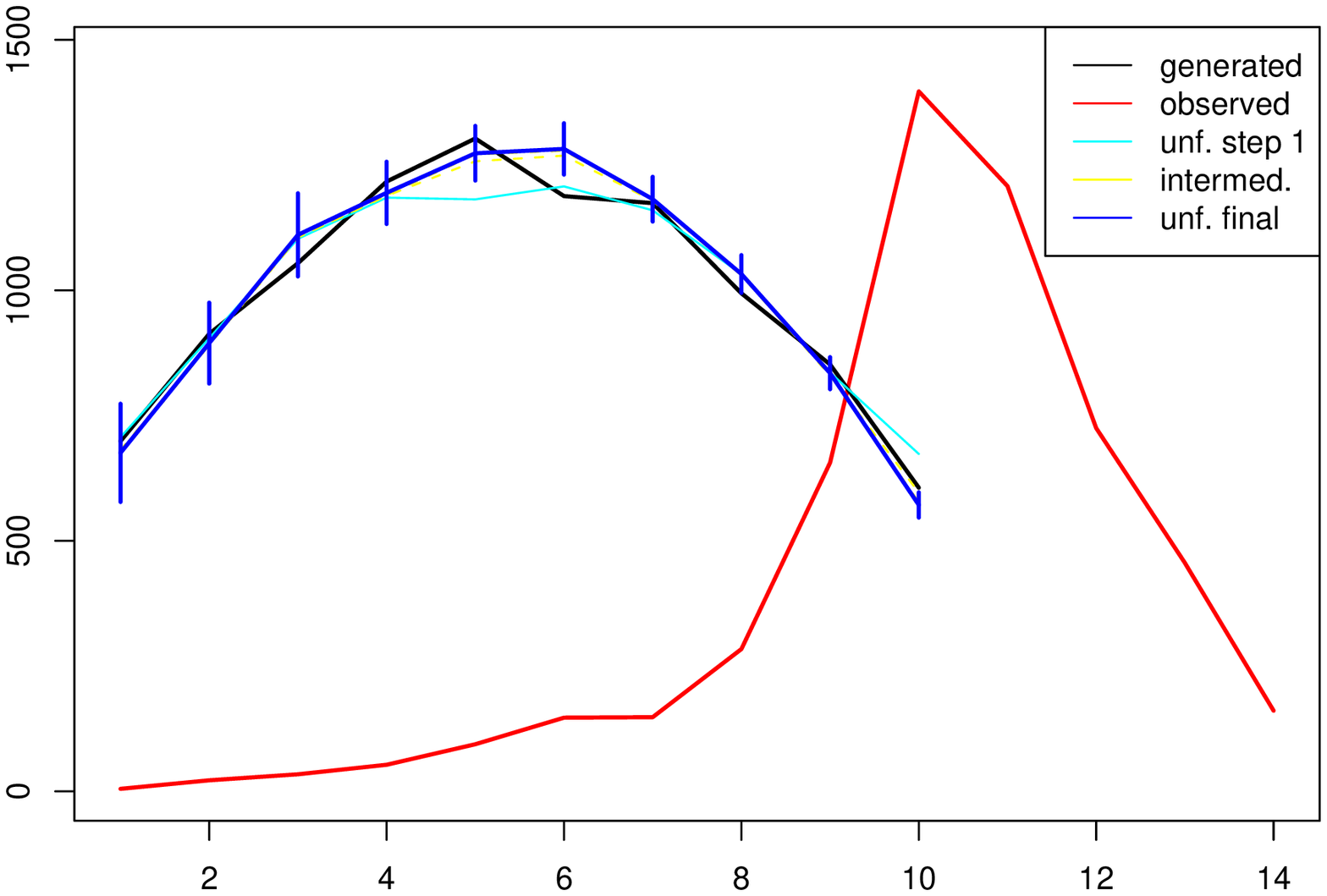,bb=40 240 555 590,clip=,width=0.47\linewidth} 
\hspace{0.5cm}
\epsfig{file=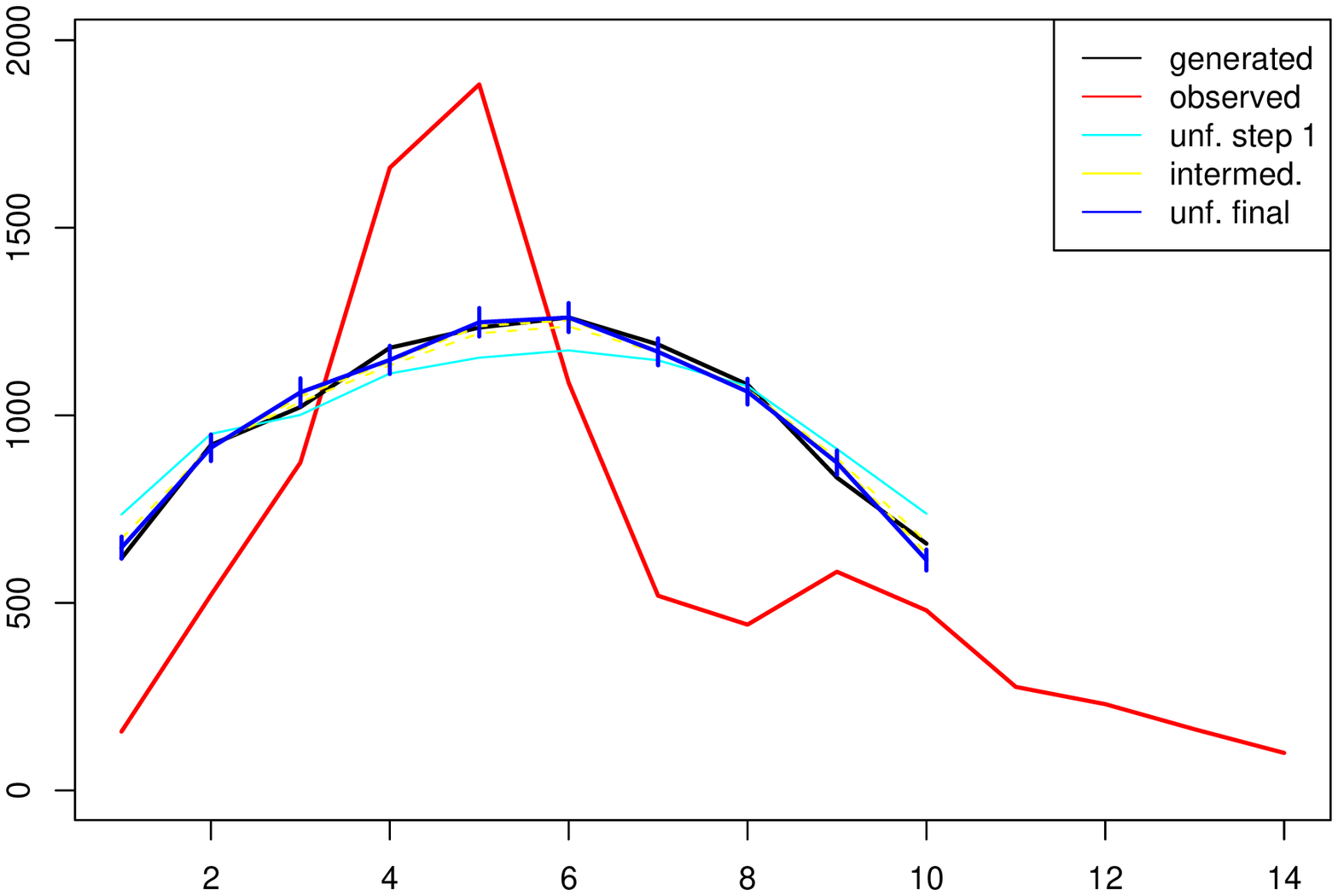,bb=40 240 555 590,clip=,width=0.47\linewidth} 
\end{center}
\caption{{\sl Same as Fig.~\ref{fig:toy1}. Independent complete simulation.}}
\label{fig:toy2}
\end{figure}

\begin{figure}
\begin{center}
\epsfig{file=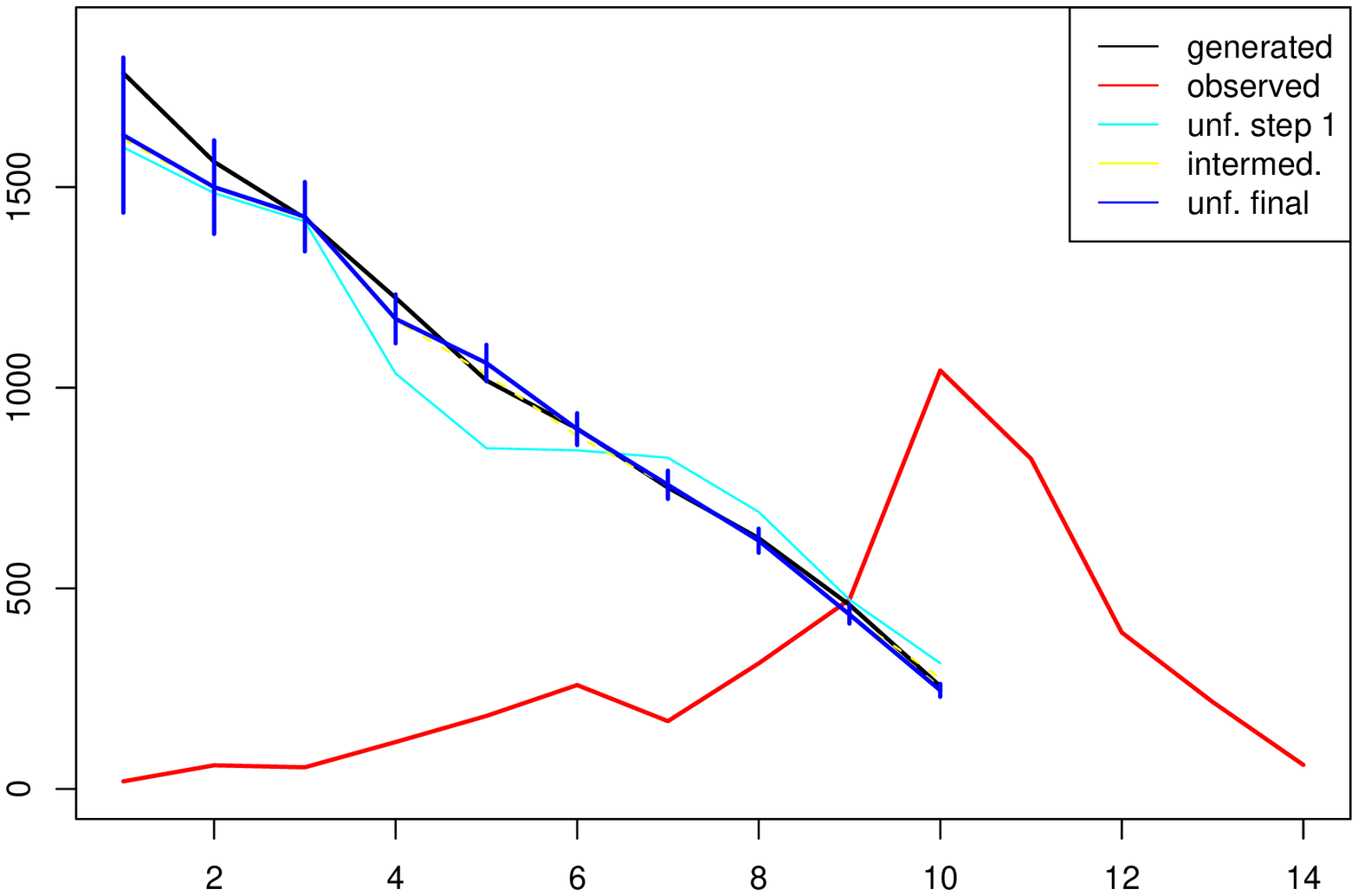,bb=40 240 555 590,clip=,width=0.47\linewidth} 
\hspace{0.5cm}
\epsfig{file=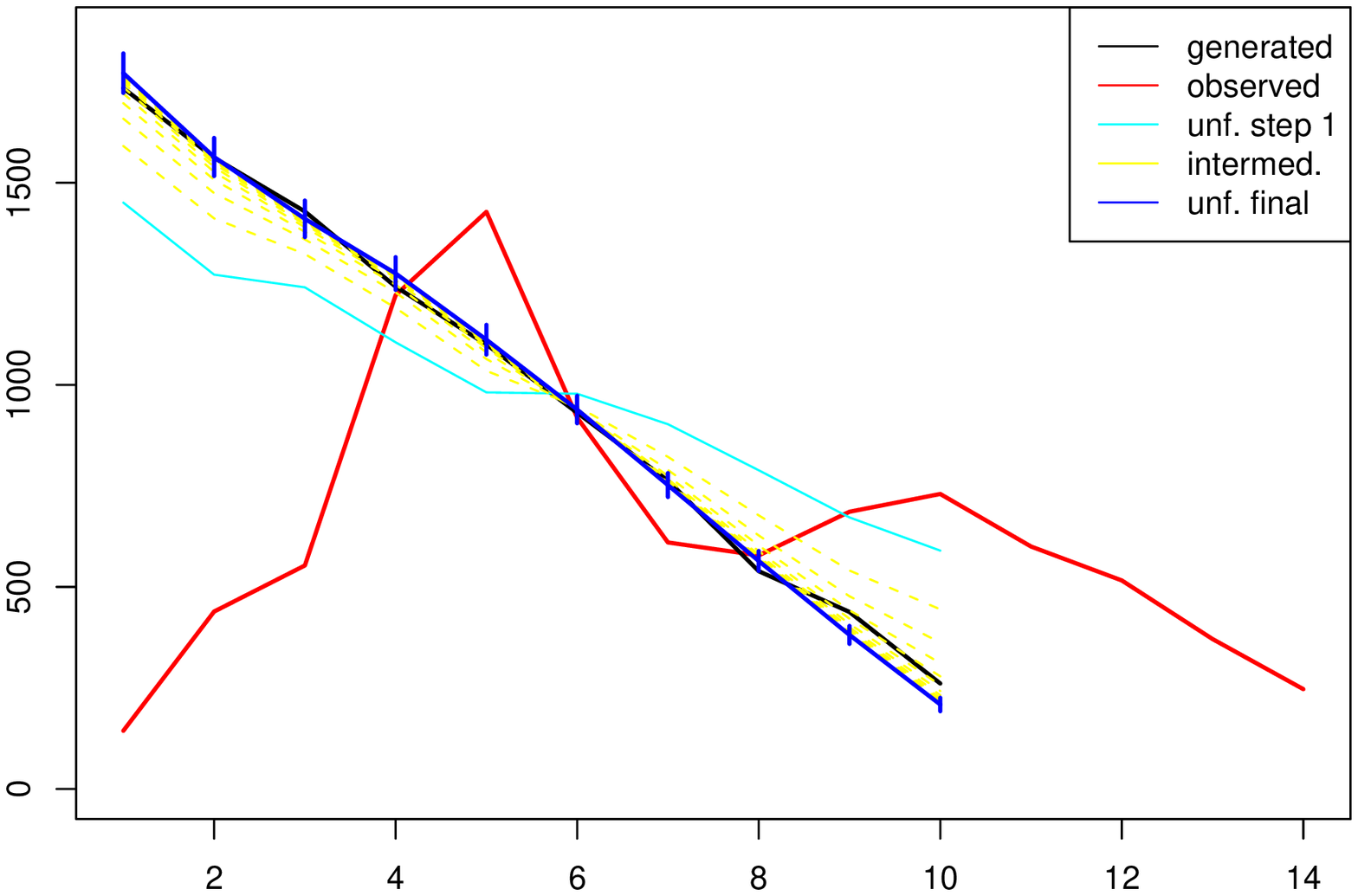,bb=40 240 555 590,clip=,width=0.47\linewidth} \\
\epsfig{file=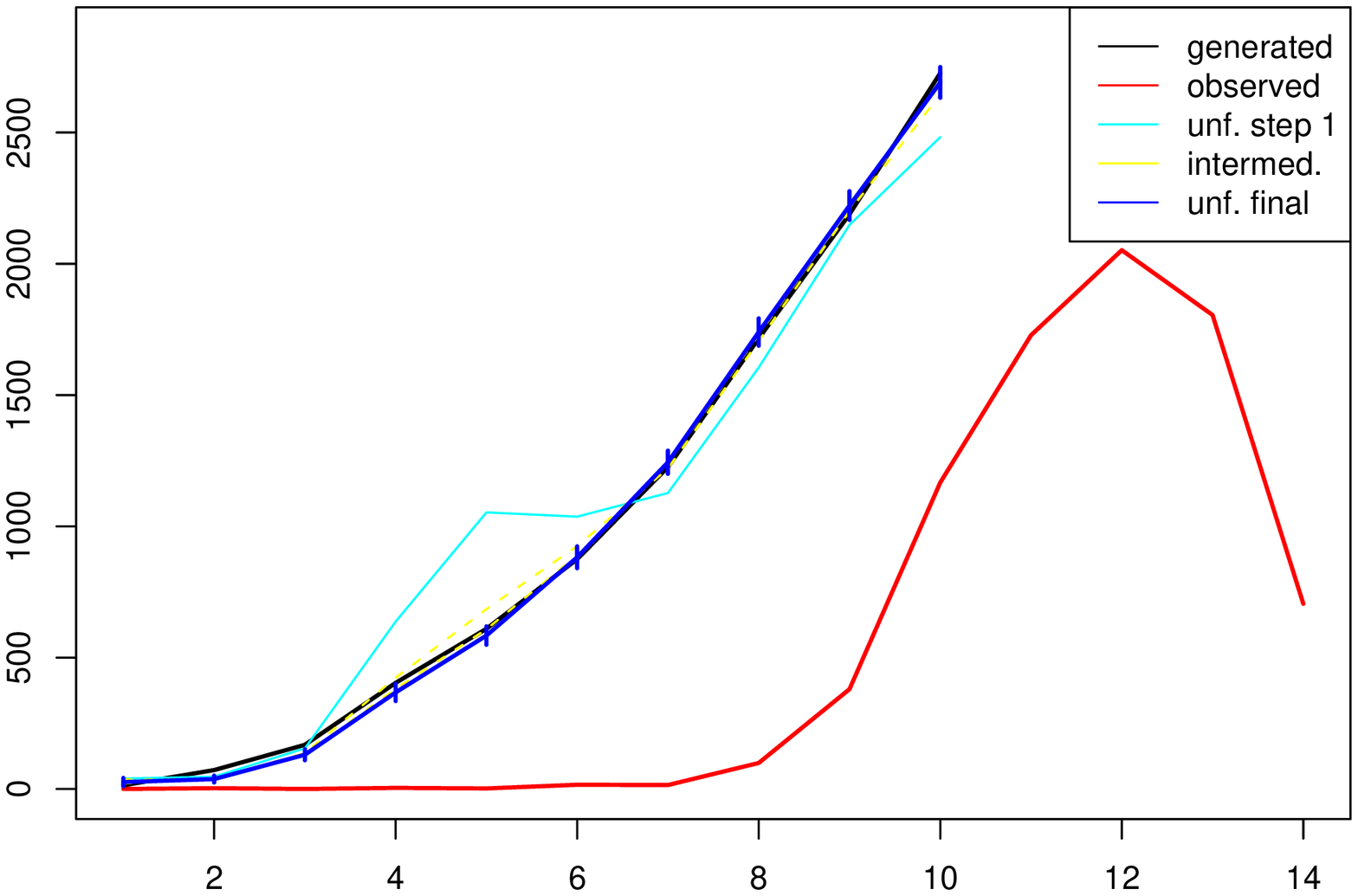,bb=40 240 555 590,clip=,width=0.47\linewidth} 
\hspace{0.5cm}
\epsfig{file=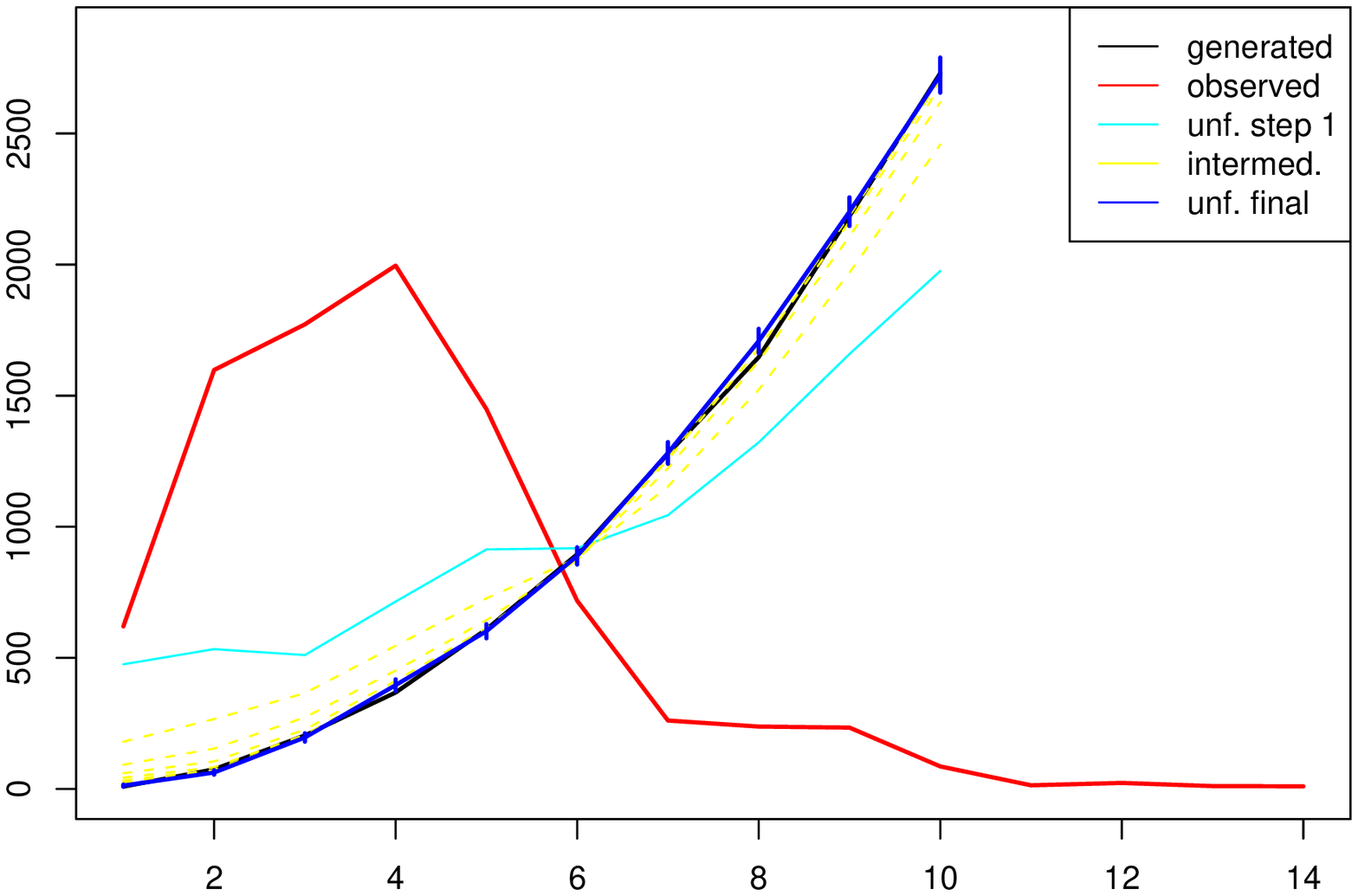,bb=40 240 555 590,clip=,width=0.47\linewidth} \\
\epsfig{file=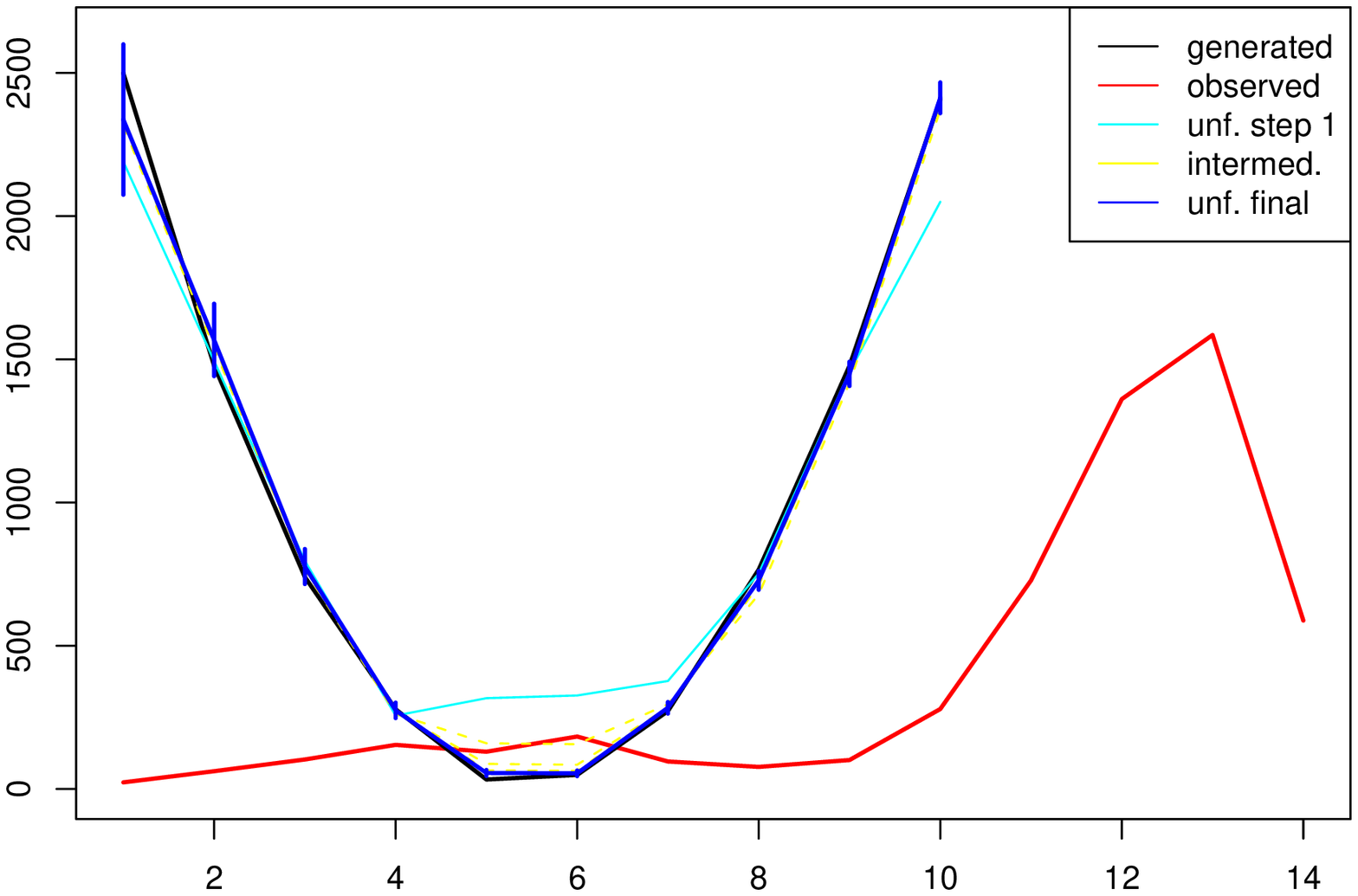,bb=40 240 555 590,clip=,width=0.47\linewidth}
\hspace{0.5cm}
\epsfig{file=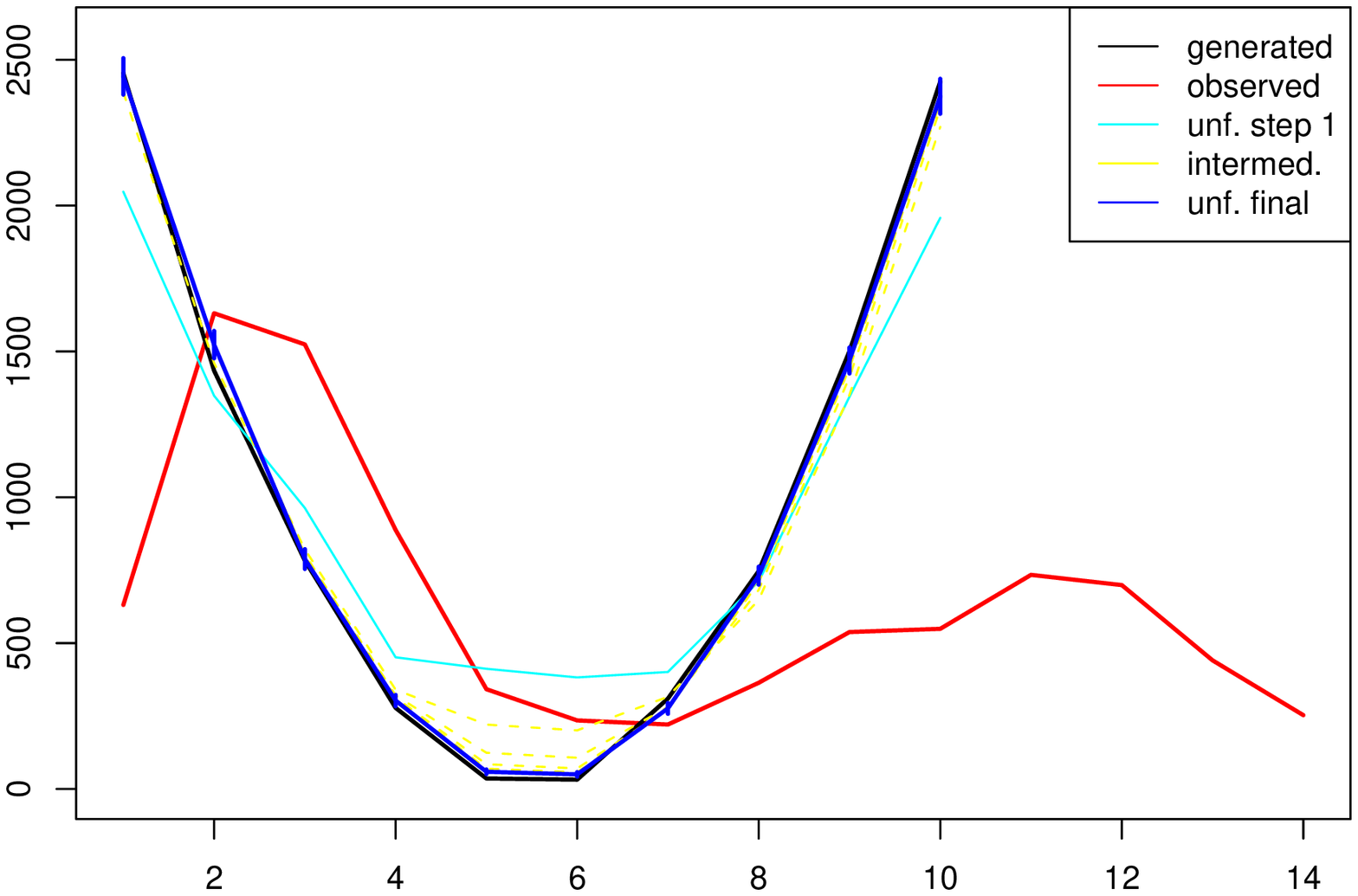,bb=40 240 555 590,clip=,width=0.47\linewidth} \\
\epsfig{file=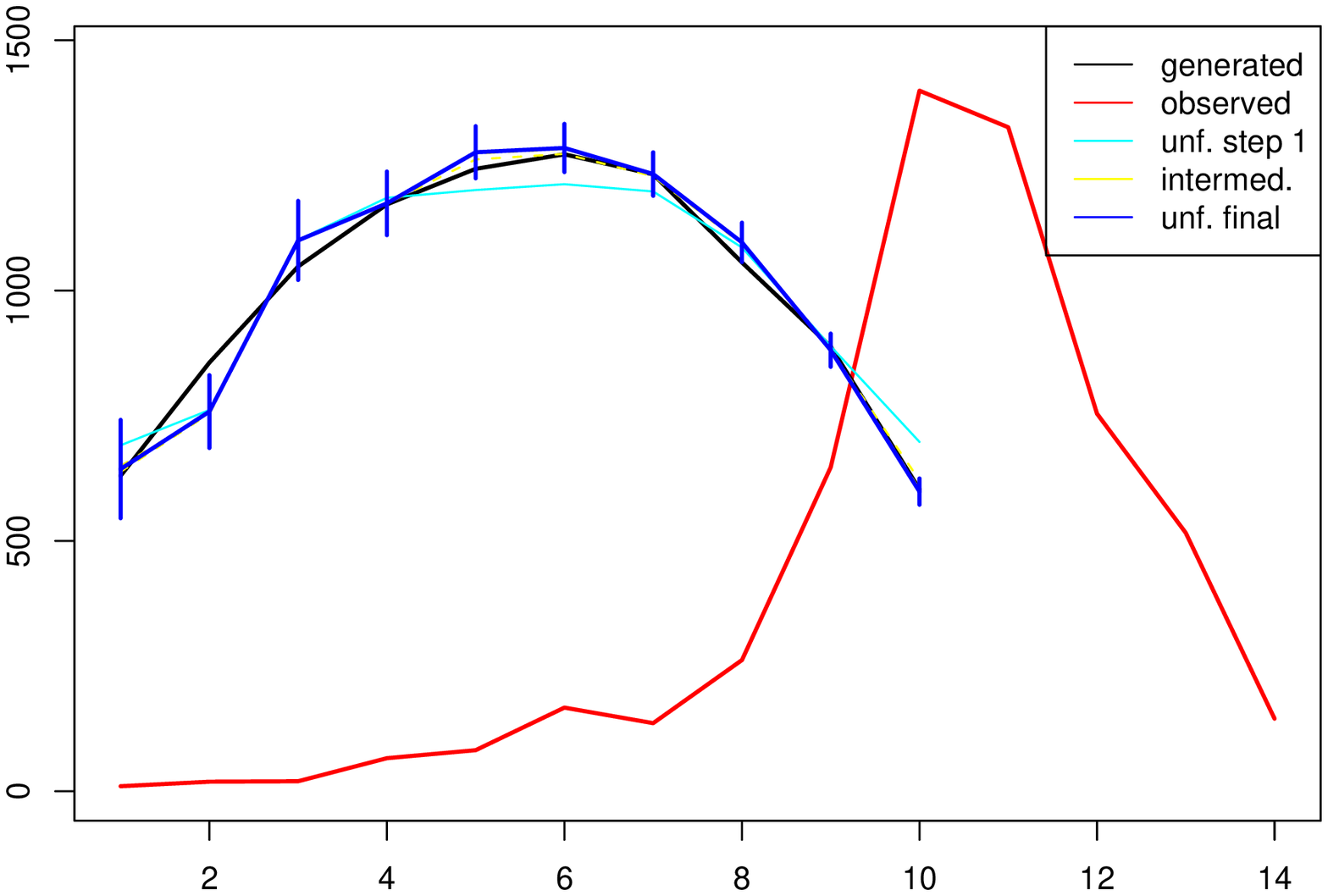,bb=40 240 555 590,clip=,width=0.47\linewidth}
\hspace{0.5cm}
\epsfig{file=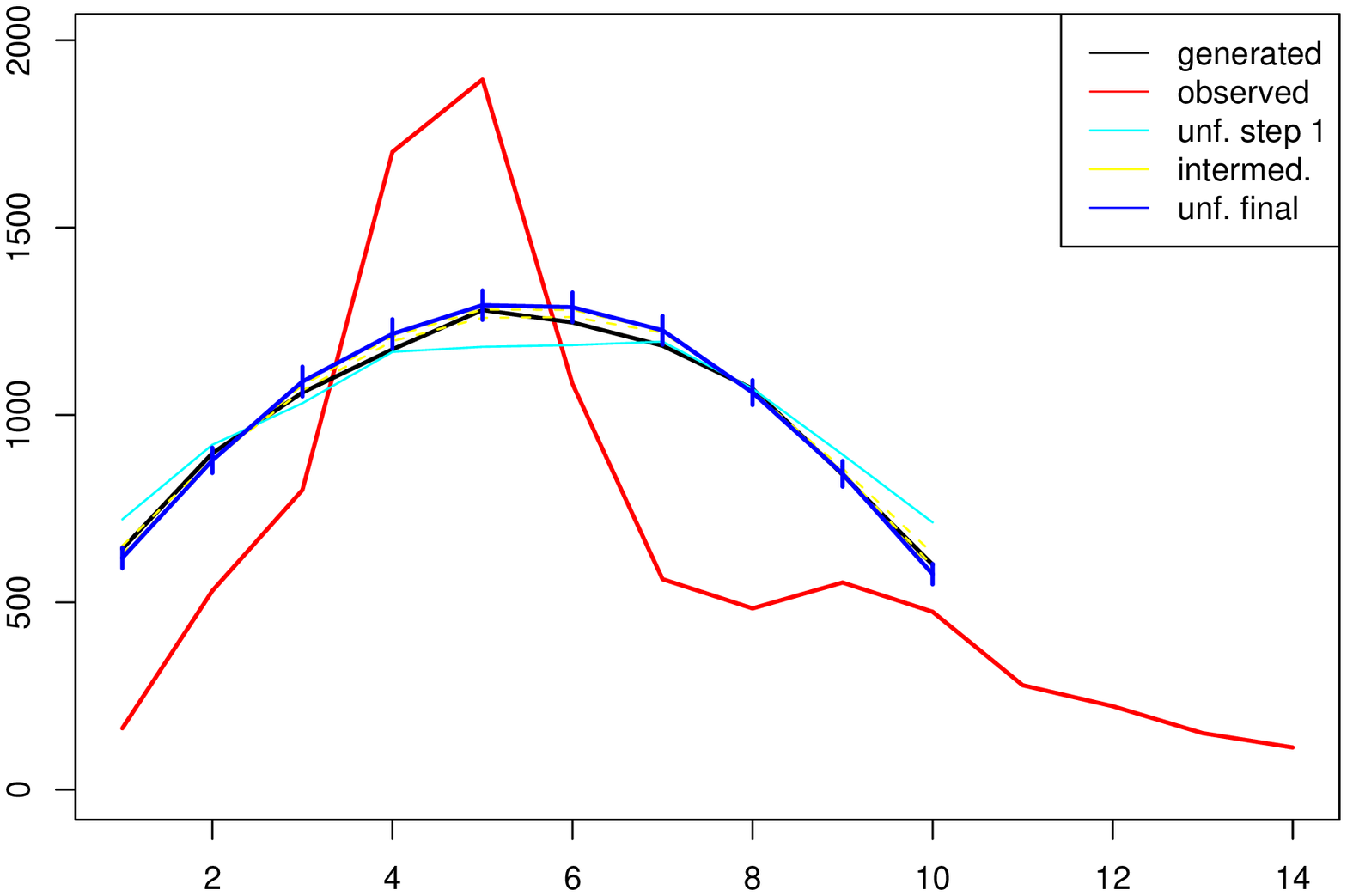,bb=40 240 555 590,clip=,width=0.47\linewidth} 
\end{center}
\caption{{\sl  Same as Fig.~\ref{fig:toy1}. Independent complete simulation.}}
\label{fig:toy3}
\end{figure}

Other independent simulations are reported in Figs.~\ref{fig:toy2} and 
\ref{fig:toy3}.

\section{Conclusions}
The evaluation of uncertainties of 
the Bayesian unfolding of Ref.~\cite{BU} has been improved
taking the probability density functions of the quantities
of interest. Simplifications are achieved using prior conjugates
and performing the relevant integrations, needed to propagate 
uncertainties, by sampling. 

The paper also discusses the issue of the iterations and the
role of the {\it intermediate} smoothing 
to reach fast convergence. It is important to note that, 
contrary to other algorithms,
the intermediate smoothing acts as a {\it regularization on the 
priors} and not on the unfolded spectrum. For this reason
physical peaks should still appear in the unfolded spectrum.

The R code of the algorithm is available, together with 
a little simulation script to run it on toy models. 
This script also implements 
of simple intermediate prior regularization, but 
the reader should have clear in mind that
this task is left to the user, who is supposed to
understand well his/her physics case.

\newpage
\section*{Appendix A -- Multinomial and Dirichlet distributions}
\subsection*{A.1\ \  Multinomial distribution}
The Multinomial distribution is the extension of the Binomial
from 2 to $k$ possible outcomes. If we assign the probability $p_i$
to the $i$-th outcome
and think of $n$ trials, then 
we are interested in the joint probability to observe 
$x_1$ times the outcome 1,  $x_2$ times the outcome 2, and so on.
From basic rules of probability theory and some
combinatorics we get
 \begin{eqnarray}
f[\,\mvec{x}\,|\,\mbox{Mult}(n,\,\mvec{p})\,] &=& 
\frac{n!}{x_1!\,x_2!\cdots x_k!}\,
p_1^{x_1}p_2^{x_2}\cdots p_k^{x_k}\,, \\
\mbox{with}\hspace{2.0cm} && \nonumber \\
\mvec{x} &=& \{x_1,x_2,\, \ldots,\,  x_k\} 
\hspace{0.90cm} [\,x_i=0,1,\ldots,n; 
             \ \  \sum_{i} x_i = n\,] \nonumber \\
\mvec{p} &=& \{p_1,p_2,\, \ldots,\,  p_k\}  
\hspace{1cm} [\,0<p<1; \ \ \sum_{i} p_i = 1\,]\,. \nonumber 
\end{eqnarray}
It is easy to see that the binomial distribution is recovered for $k=2$. 
Note that, given the constraint $\sum_ix_i=n$, 
the multinomial has to be seen as
a distribution for $k-1$ variables  $x_1,x_2,\ldots,x_{k-1}$, 
as it is clear from the binomial case. Expected values, variances,
covariances and correlation coefficients are given by
\begin{eqnarray}
\mbox{E}(X_i) & = & n\,p_i\\
\mbox{Var}(X_i) & = & n\,p_i\,(1-p_i)\\
\mbox{Cov}(X_i,X_j) & = & -n\,p_i\,p_j \hspace{3.6cm}[i\ne j] \\
\rho(X_i,X_j) &=& \frac{-p_i\,p_j}{\sqrt{p_i\,(1-p_i)\,p_j\,(1-p_j)}}
\label{eq:rho_pipj}
\hspace{1cm}[i\ne j]\,.
\end{eqnarray}
Each marginal, i.e. $f[x_i\,|\,\mbox{Mult}(n,\,\mvec{p})] $,
 is a binomial with parameters $n$ and $p_i$:
\begin{eqnarray}
f[x_i\,|\,\mbox{Mult}(n,\,\mvec{p})] &=& \frac{n!}{x_i!\,(n-x_i)}\,
p_i^{x_i}\,(1-p_i)^{n-x_i} \,.
\end{eqnarray}
For $k=2$ the two variables $X_1$ and $X_2$ 
are 100\% anti-correlated, as it is quite obvious
(the knowledge of $X_1$ determines $X_2$, and  vice versa).
In fact, since $p_2=1-p_1$, we get from Eq.~(\ref{eq:rho_pipj})
\begin{eqnarray}
\rho(X_1,X_2\,|\,k=2) &=& 
\frac{-p_1\,(1-p_1)}{\sqrt{p_1\,(1-p_1)\,(1-p_1)\,p_1}} 
= -1 \,.
\end{eqnarray}
\subsection*{A.2\ \ Dirichlet distribution}
In this case we start from the formal definition. 
Given $k$ continuous
variables whose value can range between zero and one and that sum up 
to unity, 
the Dirichlet distribution is defined by the following joint pdf:
 \begin{eqnarray}
f[\,\mvec{x}\,|\,\mbox{Dir}(\mvec{\alpha})\,] &=& 
\frac{\Gamma(\alpha_1+\alpha_2+\cdots+\alpha_k)}
     {\Gamma(\alpha_1)\,\Gamma(\alpha_2)\,\cdots\,\Gamma(\alpha_k)}\,
     x_1^{\alpha_1-1}\, x_2^{\alpha_2-1}\,\ldots \, 
      x_k^{\alpha_k-1} \hspace{0.8cm} \\
\mbox{with}\hspace{2.0cm} && \nonumber \\
\mvec{x} &=& \{x_1,x_2,\, \ldots,\,  x_k\} 
\hspace{1cm} [\,0 < x_i < 1; 
             \ \  \sum_{i} x_i = 1\,] \nonumber \\
\mvec{\alpha} &=& \{\alpha_1,\alpha_2,\, \ldots,\,  \alpha_k\}  
\hspace{1cm} [\,\alpha_i > 0\,]\,, \nonumber 
\end{eqnarray}
where $\Gamma(\,)$ is the usual gamma function.
If $k=2$, then the Beta is recovered.
Also in this case there is a constraint on the variables,
 $\sum_{i=1}^k x_i = 1$, that 
makes the distribution in fact $(k-1)$-dimensional. 
Expected values, variances and covariances are
\begin{eqnarray}
\mbox{E}(X_i) & = & \frac{\alpha_i}{\alpha}\label{eq:E_X_i_Dir}\\
\mbox{Var}(X_i) & = & \frac{\alpha_i\,(\alpha-\alpha_i)}
                           {\alpha^2\,(\alpha+1)}\label{eq:Var_X_i_Dir}\\
\mbox{Cov}(X_i,X_j) & = & \frac{\alpha_i\,\alpha_j} 
                          {\alpha^2\,(\alpha+1)}
\hspace{1cm}[i\ne j]\,,
\end{eqnarray}
where $\alpha=\sum_i\alpha_i$.  
Each marginal is a Beta with parameters $r=\alpha_i$ 
and $s=\alpha-\alpha_i$:
\begin{eqnarray}
f[x_i\,|\,\mbox{Beta}(r=\alpha_i,s=\alpha-\alpha_i)] &=& 
 \frac{x_i^{\alpha_i-1}(1-x_i)^{\alpha-\alpha_i-1}}
      {\beta(\alpha_i,\,\alpha-\alpha_i)}, 
\label{eq:dir_beta} 
\end{eqnarray}
where 
$\beta()$ is the beta function (hence the name
to the distribution), defined as 
\begin{eqnarray}
\beta(r,s)=\int_0^1 z^{r-1}(1-z)^{s-1}\,\mbox{d}z 
\ \left[=\frac{\Gamma(r)\,\Gamma(s)}{\Gamma(r+s)}\right]\,.
\end{eqnarray}
Note that the Beta pdf vanishes for $x_i=0$ if $r>1$,
and for  $x_i=1$ if $s>1$, respectively. This observation 
will be important for some considerations concerning the treatment
of bins with zero counts (see Appendix B.1).
\subsection*{A.3\ \ Dirichlet as prior conjugate of the multinomial}
If the data sample $\mvec x$, modelled by a multinomial
distribution, has been observed in $n$ trials
and we are interested in inferring the multinomial parameters
$\mvec p$, applying Bayes' theorem we get 
\begin{eqnarray}
f[\,\mvec p\,|\,\mbox{Mult}(n),\, \mvec x\,] & \propto & 
f[\,\mvec{x}\,|\,\mbox{Mult}(n,\,\mvec{p})\,]\,\cdot\, f(\mvec p) \\
& \propto & p_1^{x_1}p_2^{x_2}\cdots p_k^{x_k}\,\cdot\,f(\mvec p)\,,
\end{eqnarray}
having absorbed in the normalization all factors not 
depending on $\mvec p$.  If we model the prior $f(\mvec p)$
with a Dirichlet, i.e. 
 $f(\mvec p) \propto  p_1^{\alpha_1-1}\, p_2^{\alpha_2-1}\,\ldots \, 
      p_k^{\alpha_k-1}$, we get 
\begin{eqnarray}
f[\,\mvec p\,|\,\mbox{Mult}(n,\, \mvec x)\,] 
& \propto & (\,p_1^{x_1}p_2^{x_2}\cdots p_k^{x_k}\,)\,\cdot\, 
 (\,p_1^{\alpha_1-1}\, p_2^{\alpha_2-1}\,\ldots \, p_k^{\alpha_k-1}\,) \\
& \propto & 
 p_1^{\alpha_1+x_1-1}\, p_2^{\alpha_2+x_2-1}\,\ldots \, 
      p_k^{\alpha_k+x_k-1}\,,
\end{eqnarray}
that is still a Dirichlet. In practice,
the observation $\mvec x$ updates the Dirichlet parameters
according to the rule
\begin{eqnarray}
\mvec\alpha_{\mbox{\it posterior}} 
& = &   \mvec\alpha_{\mbox{\it prior}} + \mvec x\,. 
\label{eq:dir_alpha_update}
\end{eqnarray}
For this reason the Dirichlet is known as the {\it prior conjugate} of
the multinomial. In the simplest case of $k=2$ we have
the Beta prior conjugate of the Binomial (see e.g. Ref.~\cite{BR}).  
In particular, we can easily see that a flat prior is recovered
if all $\alpha$'s of the Dirichlet prior are equal to one.

\section*{Appendix B -- Handling the zeros}
Bins with null counts might occur either 
in the Monte Carlo results used to 
infer the smearing matrix, 
or in the observed spectrum.
As it is true that when we deal with large numbers
we can neglect fluctuations and uncertainties, 
while small numbers
are somewhat problematic, it is also true that 
handling zeros is particularly
challenging and a physicist should mistrust abstract 
mathematical arguments of any kind. 
In fact, when dealing with zeros, subjective prior
knowledge  becomes crucial,
because we need to have 
an idea of `what a zero might mean':
for some bins we could reasonably think that, if we would repeat
the same experiment or the same simulation, the zero 
might turn 
into one count, or even in two o three;
for other bins, instead, we are quite
confident that we would need to repeat the 
experiment or the simulation  many many times 
before a zero turns into one count, or perhaps 
we have good reasons to believe this will never happen. 
In other words, experienced physicists think that `not all
zeros are the same'. 

Take, for example, an
histogram in which, after some bins with positive
(and perhaps quite large)
counts, suddenly there
is a row of empty bins 
(like the second row of Tab.~\ref{tab:alphas}).
Intuitively we do not 
think all the empty bins are equivalent, although for a mathematician 
their are. But we are physicists, and we have formed
some opinion about how nature behaves. 
As a consequence, we do not only tend to smooth the positive 
numbers of the histogram, but we also 
tend to make some extrapolations
to the region of zero counts, and judge the far zeros 
to be `more zero'
that the near ones.
We need then to model somehow our opinion based on past experience.

In the following subsection we shall see, in 
a one-dimensional case, how it is
possible to treat in a consistent way
zeros that occur in the MC simulation 
needed to evaluate the smearing matrix,
where it might be really relevant.\footnote{Indeed, as it
it is common experience, `row' smearing matrices 
(i.e. not yet expressed in terms of 
Dirichlet parameters)
evaluated from MC simulations have 
many bins with zero counts (in most `sane'
situations most entries are null). Instead, there is not
such a similar problem in the observed spectrum, where the 
best pragmatic solution is re-binning, in order to
avoid zeros and even very small numbers. 
However, it is possible to imagine cases in which 
one is interested to keep one or a few bins separated from
the others, even no counts have been observed in it/them. 
These special bins can be treated {\it ad hoc}, in a way
similar to that discussed in the following subsection, 
playing on the parameters of the relevant gamma prior.} 
Indeed, as discussed in section \ref{sec:iter}
about smoothing, this is a task left to the
user, which knows the physical meaning of the 
bins.\footnote{But don't worry: the program gives you the option 
of treat the zeros as \ldots \,zeros, and you might want to skip
the following two subsections (that might be a bit `academic'
and of little practical relevance in most cases).
But you might want to reflect a while that, perhaps, considering,
democratically, all zeros on equal foot
might be against your beliefs, while probably you do not want to be 
incoherent\ldots}

\subsection*{B.1 -- Zero counts in effect-bins of MC events 
used to infer smearing matrix}\label{ss:zero_MC}
 When we make Monte Carlo
simulations in order to infer the smearing matrix we 
usually observe many zeros, because long range migrations
are usually rather rare in typical 
detectors.\footnote{Nevertheless, when we
are interested in reconstructing physics quantities computed
on an ensemble of tracks and clusters, and the detector
has not negligible inefficiencies, the migrations can be huge, has 
it was used to happen in 'two-photon physics' in e$^+$-e$^-$
colliders, or in some kinematical regions of deep inelastic scattering,
as impressively shown in Fig.  6.8  of page 77 of Ref.~\cite{Nakao} 
(the choice of this reference is that similar figures, 
rather popular at the beginning of HERA physics, are usually
only available on printed notes and conference proceedings, 
of difficult accessibility).}

As explained in Sec.~\ref{sec:BayesIntro}, the
columns of the smearing matrix are jointly described by
vectors of Dirichlet random variables 
(one vector per cause-bin). The parameters of a  Dirichlet
pdf are updated according to Eq.~(\ref{eq:dir_alpha_update}),
that we rewrite here:
\begin{eqnarray*}
\mvec\alpha_{\mbox{\it posterior}} 
& = &   \mvec\alpha_{\mbox{\it prior}} + \mvec x\,,
\end{eqnarray*}
where $\mvec x$ stands here for 
$\left.\mvec x_E^{MC}\right|_{x(C_i)}$
[see Eq.~(\ref{eq:update_alpha})].
A usual choice of the Dirichlet prior to model indifference
is to set all $\alpha$'s to 1. 
The prior is then characterized [see Eqs.\,(\ref{eq:dir_beta}),
 (\ref{eq:E_X_i_Dir}) and  (\ref{eq:Var_X_i_Dir})] 
by the the following marginals, 
expectations and variances (we use the symbol $p$ for the variables, 
for their physical meaning in our contest, and the index $j$ for numbering 
them, as we do in the text for the effect-bins):
\begin{eqnarray}
f[p_j\,|\,\mbox{Beta}(1,\, \nu-1)] &=& 
\frac{(1-p_j)^{\nu-2}}{\beta(1,\,\nu-1)}\\
\mbox{E}(p_j) & = & \frac{1}{\nu}\\
\mbox{Var}(p_j) & = &\frac{\nu-1}{\nu^2\,(\nu+1)} 
              \xrightarrow [\ \ \nu \gg 1\ \ ]{}{\frac{1}{\nu^2}}\,,
\end{eqnarray}
where, hereafter, 
$\nu$ is used in place of  $n_E+1$ 
in order to improve the readability of the formulas.
In the case of $\nu=2$ we recover a flat prior and hence an
expectation of $1/2$. 

If we throw $n$ MC events in order to estimate the several $p_j$,
the updated $\alpha$ parameters become $\alpha_j=x_j+1$.
For the marginals we have now:
\begin{eqnarray}
f[p_j\,|\,\mbox{Beta}(x_j+1,\, n+\nu-x_j-1)] 
\!&=&\! \frac{p^{x_j}\,(1-p_j)^{n+\nu-x_j-2}}{\beta(x_j+1,\,n+\nu-x_j-1)}\,,
\hspace{0.5cm}
\end{eqnarray}
with expected values and variances
\begin{eqnarray}
\mbox{E}(p_j) \!& =\! & \frac{x_j+1}{n+\nu}\\
\mbox{Var}(p_j) \!& = &\!\frac{(x_j+1)\,(n+\nu-x_j-1)}{(n+\nu)^2\,(n+\nu+1)}\,.
\end{eqnarray}
In the limit of large numbers ($n\gg \nu$ and $x_j\gg 1$) 
we get an expected value equal to $x_j/n$ (that is 
the fraction of events in the cell $j$, in agreement
with a naive evaluation of the elements of the
smearing matrix), with a variance of 
$(x_j/n)\,(1-x_j/n)\,/\,n$. 

In the case {\it zero counts 
and large $n$} the pdf is given by
\begin{eqnarray}
f(p_j\,|\,x_j=0) = (n+1)\, (1-p_j)^{n}\,,
\label{eq:beta_x0}
\end{eqnarray}
having a sharp peak at $p_j=0$, with sharpness growing 
with $n$. Expected number and standard
deviation are  $\mbox{E}[p_j\,|\,x_j]=\sigma[p_j\,|\,x_j]=1/n$
[moreover, in this case the cumulative has the easy expression
$F(p_j\,|\,x_j=0)=1-(1-p_j)^{n+1}$, from which we can calculate a 
95\% probability upper limit for $p_j$ to be 
$1-0.05^{1/(n+1)}$ 
that is approximately $3/n$ for large $n$].
\subsubsection*{Solving a formal paradox with some good sense}
The above conclusions seem reasonable when
 applied to an individual empty bin.
But let us see, with the help of Tab.~\ref{tab:alphas},
 what happens where several elements of $\mvec x$ are null:
\begin{table}
{\footnotesize
\begin{center}
\begin{tabular}{|l|c|c|c|c|c|c|c|c|c|c|c|c|}
\hline
$E_j$   &  1 & 2 & 3 & 4 & 5 & 6 & 7 & 8 & 9 & 10 & 11 & 12  \\ 
\hline
$x(E_j)$ & 10 & 20 & 40 & 15 & 10 & 5 & 0 & 0 & 0 & 0 & 0 & 0  \\
\hline
& \multicolumn{12}{|c|}{Standard uniform priors} \\
\hline 
$\alpha_{pr_j}$ & 1 & 1 & 1 & 1 & 1 & 1 &  1 & 1 & 1 & 1 & 1 & 1  \\
$\alpha_{post_j}$ &11 & 21 & 41 & 16 & 11 & 6 & 1 & 1 & 1 & 1 & 1 & 1  \\
$\mbox{E}[p_{j}]$ 
& 9.8 & 18.8 & 36.6 & 14.3 & 9.8 & 5.4 & 
0.9 & 0.9 & 0.9 & 0.9 & 0.9 & 0.9   \\
$\sigma[p_{j}]$ 
& 2.8 & 3.7 & 4.5 & 3.3 & 2.8 & 2.1 & 
0.9 & 0.9 & 0.9 & 0.9 & 0.9 & 0.9   \\
\hline 
 & \multicolumn{6}{|c|}{$\begin{array}{l}\mbox{E}[p_{1-6}]=94.6\% \\   
                          \sigma[p_{1-6}]=2.1\%\end{array}$}  & 
\multicolumn{6}{|c|}{$\begin{array}{l}\mbox{E}[p_{7-12}]=5.4\% \\   
                          \sigma[p_{7-12}]=2.1\%\end{array}$} \\
\hline 
& \multicolumn{12}{|c|}{$\alpha_j=2/\nu$} \\
\hline 
$\alpha_{pr_j}$ & 1/6 & 1/6 & 1/6 & 1/6 & 1/6 & 1/6 &  
                  1/6 & 1/6 & 1/6 & 1/6 & 1/6 & 1/6  \\
$\alpha_{post_j}$ &10.2 & 20.2 & 40.2 & 15.2 & 10.2 & 5.2 & 
                   0.2 & 0.2 & 0.2 & 0.2 & 0.2 & 0.2  \\
$\mbox{E}[p_{j}]$ 
& 10.0 & 19.8 & 39.4 & 14.9 & 10.0 & 5.1 & 
0.02 & 0.02 & 0.02 & 0.02 & 0.02 & 0.02   \\
$\sigma[p_{j}]$ 
& 3.0 & 3.9 & 4.8 & 3.5 & 3.0 & 2.2 & 
0.4 & 0.4 & 0.4 & 0.4 & 0.4 & 0.4   \\
\hline 
 & \multicolumn{6}{|c|}{$\begin{array}{l}\mbox{E}[p_{1-6}]=99.0\% \\   
                          \sigma[p_{1-6}]=1.0\%\end{array}$}  & 
\multicolumn{6}{|c|}{$\begin{array}{l}\mbox{E}[p_{7-12}]=1.0\% \\   
                          \sigma[p_{7-12}]=1.0\%\end{array}$} \\
\hline 
& \multicolumn{12}{|c|}{Regularized $\alpha$'s in bins with zero counts} \\
\hline 
$\alpha_{pr_j}$ & 1/6 & 1/6 & 1/6 & 1/6 & 1/6 & 1/6 &  
                  1/6 & 1/6 & 1/6 & 1/6 & 1/6 & 1/6  \\
$\alpha_{post_j}$ &10.2 & 20.2 & 40.2 & 15.2 & 10.2 & 5.2 & 
                   0.2 & 0.2 & 0.2 & 0.2 & 0.2 & 0.2  \\
$f$ & 1 & 1 & 1 & 1 & 1 & 1 & 1 & 1/2 & 1/4 & 1/8 & 1/16 & 1/32 \\  
$\alpha_{post_j}^{(r)}$ &10.2 & 20.2 & 40.2 & 15.2 & 10.2 & 5.2 & 
                   0.2 & 0.1 & 0.04 & 0.02 & 0.01 & 0.005  \\
$\mbox{E}[p_{j}]$ 
& 10.0 & 19.9 & 39.6 & 15.0 & 10.0 & 
5.1 & 0.2 & 0.1 & 0.04 & 0.02 & 0.01 & 0.005 \\
$\sigma[p_{j}]$ 
& 3.0 & 3.9 & 4.8 & 3.5 & 3.0 &
2.2 & 0.4 & 0.3 & 0.2 & 0.1 & 0.1 & 0.1  \\
\hline 
 & \multicolumn{6}{|c|}{$\begin{array}{l}\mbox{E}[p_{1-6}]=99.7\% \\   
                          \sigma[p_{1-6}]=0.6\%\end{array}$}  & 
\multicolumn{6}{|c|}{$\begin{array}{l}\mbox{E}[p_{7-12}]=0.3\% \\   
                          \sigma[p_{7-12}]=0.6\%\end{array}$} \\
\hline 
& \multicolumn{12}{|c|}{Regularized and rescaled $\alpha$'s \ 
                   $[\,*:\, \sum_{k\subset{\cal M}} \alpha_k = m/\nu\,]$}\\
\hline 
$\alpha_{pr_j}$ & 1/6 & 1/6 & 1/6 & 1/6 & 1/6 & 1/6 &  
                  1/6 & 1/6 & 1/6 & 1/6 & 1/6 & 1/6  \\
$\alpha_{post_j}$ &10.2 & 20.2 & 40.2 & 15.2 & 10.2 & 5.2 & 
                   0.2 & 0.2 & 0.2 & 0.2 & 0.2 & 0.2  \\
$f$ & 1 & 1 & 1 & 1 & 1 & 1 & 1 & 1/2 & 1/4 & 1/8 & 1/16 & 1/32 \\  
$\alpha_{post_j}^{(r\&r)}\,*$ &10.2 & 20.2 & 40.2 & 15.2 & 10.2 & 5.2 & 
                   0.5 & 0.3 & 0.13 & 0.06 & 0.03 & 0.015  \\
$\mbox{E}[p_{j}]$ 
& 10.0 & 19.8 & 39.4 & 14.9 & 10.0 & 
5.1 & 0.5 & 0.2 & 0.1 & 0.06 & 0.03 & 0.02 \\
$\sigma[p_{j}]$ 
& 3.0 & 3.9 & 4.8 & 3.5 & 3.0 &
2.2 & 0.7 & 0.5 & 0.3 & 0.2 & 0.2 & 0.1  \\
\hline 
 & \multicolumn{6}{|c|}{$\begin{array}{l}\mbox{E}[p_{1-6}]=99.0\% \\   
                          \sigma[p_{1-6}]=1.0\%\end{array}$}  & 
\multicolumn{6}{|c|}{$\begin{array}{l}\mbox{E}[p_{7-12}]=1.0\% \\   
                          \sigma[p_{7-12}]=1.0\%\end{array}$} \\
\hline 
\end{tabular}
\end{center}
}
\caption{{\sl MC events, Dirichlet parameters 
$\mvec \alpha$ and expectations of
 Dirichlet variables $p_j$ (value in \%) with  several treatments of the zeros
(see text).}}
\label{tab:alphas}
\end{table}
one hundred events have been generated in a  cause-cell
and the second row of Tab.~\ref{tab:alphas}, 
whose elements are indicated 
by $x(E_j)$, shows the counts in each effect-bin. 
The third row shows the prior  $\mvec \alpha$ and 
the forth one the updated $\mvec \alpha$.
We also calculate in the following two rows 
expected values and standard
deviation of the $p_j$ values.

All elements of the smearing matrix that connect that cause-cell
 the effect-cells with zero counts have  
$\mbox{E}[p_j]=\sigma[p_j]\approx 0.9\%$. 
This looks reasonable if we consider each cell individually. 
But, observing the values of $\mbox{E}[p_j]$ and $\sigma[p_j]$
in the fifth and sixth row of Tab.~\ref{tab:alphas},
 any physicist would be uneasy, for a couple 
of reasons. First, we are have strong prejudices towards regularity
of physical laws, including whatever causes the smearing. 
If the smearing is peaked around bin $E_3$ and drops in both sides, 
we do not expect the smearing probability $p_{12}=P(E_{12}\,|\,C_i\,,I)$ 
to be equal to $p_7=P(E_{7}\,|\,C_i\,,I)$. 
Second, if we sum up the probabilities
of all bins with no counts, i.e. $\sum_{j=7}^{12}\mbox{E}[p_j]$,
 the total is not really negligible. 
The situation becomes worse if we add more 
bins in the right side. 

Another way to understand the problem is starting from the 
probability density function of $p_{\cal M}$, the parameter
of a binomial distribution that describes the occurrence
of an event in the set ${\cal M}$ of 
$m$ selected bins
(not necessarily adjacent, although in Tab.~\ref{tab:alphas} they are). 
Just extending 
a well known property of the Dirichlet distribution
[see Appendix A.2, and in particular Eq.~(\ref{eq:dir_beta})],
we get that $p_{\cal M}$ is described by a Beta pdf, 
with parameter $\alpha_{\cal M} = \sum_{k\subset {\cal M}} \alpha_k$
and $\alpha-\alpha_{\cal M}$, where $\alpha=\sum_j\alpha_j$:
\begin{eqnarray}
f[p_{\cal M}\,|\,\mbox{Beta}(\alpha_{\cal M},\,\alpha-\alpha_{\cal M})] &=& 
 \frac{p_{\cal M}^{\alpha_{\cal M}-1}
       (1-p_{\cal M})^{\alpha-\alpha_{\cal M}-1}}
      {\beta(\alpha_{\cal M},\,\alpha-\alpha_{\cal M})}\,. 
\label{eq:dir_beta_M}
\end{eqnarray}
If we take ${\cal M}$ to be the set of the $m$ 
bins with zero counts ($m=6$ in the case of Tab.~\ref{tab:alphas})
and for each of them we choose the initial
value of $\alpha$ to be equal 1,
we get:
\begin{eqnarray}
f[p_{\cal M}\,|\,\mbox{Beta}(m,\, n - m)] &=& 
 \frac{p_{\cal M}^{m-1}
       (1-p_{\cal M})^{n-m-1}}
      {\beta(m,\,n-m)}\,. 
\label{eq:dir_beta_M_zero}
\end{eqnarray}
Contrary to Eq.~(\ref{eq:beta_x0}), this pdf vanishes for 
$p_{\cal M} \rightarrow 0$, 
thus stating {\it we are a priori 
sure that $p_{\cal M}$ cannot be zero}. This is a bit too much! 
We understand, then, that the required condition 
in order to have $f(p_{\cal M}=0) = 0$ is to apply the following 
constraint to the $\alpha$'s:
$\sum_{k\subset {\cal M}} \alpha_k\le 1$,
that becomes $\alpha_{k\subset {\cal M}} \le 1/m$ in the case we
consider all zero count bins on the same foot. 
Since the final $\alpha$ of the bins with many counts is 
influenced very little by the initial $\alpha$, if this 
does not exceed ${\cal O}(1)$, then
we could simply require $\alpha_j\approx 1/\nu$. But, just to recover
the uniform prior when $\nu=2$, which can be written
as a Dirichlet with $\alpha_1=1$ and $\alpha_2=1$ (that is  
a Beta with $r=1$ and $s=1$), our starting point 
will be to set $\alpha = 2/\nu$ for all bins. 
Then this value will be modified due to the constraint 
$\sum_{k\subset {\cal M}} \alpha_k = m/\nu$ for the 
bins with  zero counts. The result of this strategy is 
reported in Table \ref{tab:alphas} in the rows just
below ``$\alpha_j = 2/\nu$''.

\subsubsection*{Regularization of the smearing matrix prior}
There is still the problem that we do not 
believe all zeros are the same. We can solve it shaping in some 
way the $\alpha$'s in a suitable way. Obviously, in order to do this,
we need to know the bin ordering in the parameter space, that
was not taken into account till now. We show here
a simple way of how this regularization 
can be performed in a 1-dimensional spectrum unfolding.
We set all initial $\alpha$'s to $2/\nu$ in all bins 
that have non-zero counts and in the adjacent ones. Then, as we move 
far away from the bins with non-zero counts, we decrease the value 
of alpha, for example with some geometric law 
(i.e. exponential decrease). In Tab.~\ref{tab:alphas} we have done 
this exercise taking 
the geometric factor equal $1/2$ 
(see the rows indicated by `$f$'). 
That is perhaps a too conservative choice
that still takes in too much considerations the tails
(indeed, in the program the geometric factor is given as an 
option in the program, whose default value is equal to $1/e$, i.e.
a `natural' exponential decrease that gives 
roughly one order of magnitude suppression every two bins).
The $\alpha$'s shaped in this way are finally rescaled
with the condition $\sum_{k\subset {\cal M}} \alpha_k = m/\nu$.%
\break \newpage

To conclude, these are the steps made in the default 
practical method to handle zeros in 1-dimensional unfolding:
\begin{itemize}
\item
All bins: 
\begin{itemize}
\item
set $\alpha_j = 2/\nu + x_j$ in all bins ($\alpha_{pr_j}$ in the table).
\end{itemize}
\item
Only bins with zero counts:
\begin{itemize}
\item
reshape the $\alpha$'s with an exponentially  
law;\footnote{In the case of a sequence of zero counts bins 
surrounded on both sides by non-zero bins, the two exponential
starting at each side are summed up, to mimic the fact 
that there might be an incoherent sum of the two tails;}
\item
rescale the $\alpha$'s by the condition
$\sum_{k\subset {\cal M}} \alpha_k = m/\nu$\\
\mbox{}[$\alpha_{post_j}^{(r\&r)}$ in the table].
\end{itemize}
\end{itemize}
The last rows of Tab.~\ref{tab:alphas} show the results 
obtained applying this rule, which seems a reasonable compromise
between the several pieces of information and beliefs 
as they have been stated here. Needless to say, you are
free to adjust the treatments of the zero according to the best knowledge
of your physical case. The program offers you the possibility to
return your preferred $\alpha$'s using the function {\tt my.alphas()},
or just to decide that zeros are simply zeros, and {\it basta}.

\subsubsection*{A warning about the Monte Carlo checks of Bayesian methods}
A last comment about the outcomes of the
simulations performed to prove the correctness
of the above procedure is in order. In fact you might want
to check the unfolding performing a large number
of simulated experiments. Most likely you will arrive
to the conclusion that the strategy to treat the zeros discussed above
will provide in average a biased result. This is not a surprise, 
since you are setting some smearing probabilities to positive values,
whereas they are most likely exactly zero in the simulation. 
This effect is discussed in section 10.6 of Ref.~\cite{BR} and you should
pay attention to it and make the simulation in the correct way, 
before stating that ``Bayesian methods yield biased estimations''.

\subsection*{B.2 -- Zero counts in real data effect-bins}\label{ss:zero_data}
Having explained in length {\it a} strategy to treat the bins with
zero counts in the evaluation of the smearing matrix, 
one could thing of doing something similar with the empty bins
of the experimental spectrum. However it is clear that in this second
case the idea of handling in a sophisticate way the zeros might
sound purely academic, because an appropriate re-binning is in
most cases the simplest pragmatic solution. 

Anyway, since a possible strategy follows from what it
has been discussed in Appendix B.1, let us sketch it here  briefly
with the benefit of the inventory.

As we have seen in section \ref{ss:improvements}, 
the default initial $c$ and $r$ parameters of 
the gamma pdf [see Eq.~(\ref{eq:mu_j_Gamma})] are 1 and 0, 
respectively, corresponding to an {\it improper} 
flat prior (extending to infinite, with infinite
expected value). The  Bayesian updating described 
by Eq.~(\ref{eq:mu_j_Gamma})
makes the posterior (mathematically) nice (i.e. proper) 
and (physically) meaningful. 
In particular, empty bin $j$ yields 
\begin{eqnarray}
f[\mu_j\,|\,x(E_j)=0] &=& e^{-\mu_j}\,,
\end{eqnarray}
having null mode and unity expected value, that is in order.

The problem occurs when there are several empty bins,
because if we treat them all in this way, we get that 
the pdf of the sum of the respective $\mu$'s (let us
call it $\Sigma$) vanishes 
when its value goes to zero. Again, a possible 
solution is to constrain to zero the mode of $\Sigma$.
This can be done  playing with the
$c$ and $r$ parameters of the empty bins, 
as done, for example, by
the {\it custom} function 
(real cases require intelligent inputs by the user!) 
{\tt my.gamma.par()} of the distribution
code~\cite{R-code}.
[Note that, contrary to {\tt my.alphas()}, which 
tries to treat with a certain continuity 
adjacent zero bins, {\tt my.gamma.par()} 
is more primitive, because, as said above, it
is not considered important for practical applications.]

\end{document}